\documentclass[9pt,twocolumn,twoside]{opticajnl}
\journal{opticajournal}
\setboolean{shortarticle}{false}
\setboolean{displaycopyright}{false}
\dates{}
\doi{}
\usepackage{lineno}

\title{Relativistic Harmonic Generation in Spatiotemporal Plasma Photonic Crystals}
\author[1]{Haojie Zhang}
\author[1,*]{Boyuan Li}
\author[1,2]{Xiangyan An}
\author[1]{Rong Huang}
\author[1]{Feng Liu}
\author[1,2]{Zheng-Ming Sheng}
\author[1,*]{Min Chen}
\affil[1]{State Key Laboratory of Dark Matter Physics, Key Laboratory for Laser Plasmas (MoE), School of Physics and Astronomy, Shanghai Jiao Tong University, Shanghai 200240, China}
\affil[2]{Tsung-Dao Lee Institute, Shanghai Jiao Tong University, Shanghai 200240, China}
\affil[*]{minchen@sjtu.edu.cn (Min Chen) \quad liby1990@sjtu.edu.cn (Boyuan Li)}

\begin{abstract}
Spatiotemporal photonic crystals offer a route to controlling the frequency, momentum, and energy of light. Extending this capability to relativistic laser intensities calls for a plasma-based platform. Here we propose a spatiotemporal plasma photonic crystal (SPPC) formed at a solid surface driven by intense pump pulses, with a separate probe interacting with the resulting modulation. The spatial and temporal periodicity of the SPPC determines the allowed harmonic frequencies and emission angles, while its complex Floquet bands explain how the probe draws energy from the modulation. Beyond harmonic generation, this picture predicts wavelength-dependent probe amplification as the probe accesses amplifying modes of the SPPC, providing an experimentally testable connection between the band structure and energy transfer. Agreement with particle-in-cell simulations supports this interpretation.
\end{abstract}

\begin{document}
\pagestyle{plain}
\maketitle
\thispagestyle{plain}

\section{Introduction}
A spatiotemporal photonic crystal (SPC) is an optical medium whose electromagnetic properties vary periodically in both space and time \cite{Sharabi:22,Ye2025ActiveMetasurface}. It generalizes spatial photonic crystals \cite{PhysRevLett.58.2059,PhysRevLett.58.2486} and photonic time crystals \cite{Lustig:18} by combining spatial and temporal periodicity. This combined periodicity can impart both frequency and momentum shifts to incident light \cite{Sharabi:22,PhysRevE.75.046607}. Temporal periodicity also allows energy transfer between the modulation and the optical field, as illustrated in photonic time crystals \cite{doi:10.1126/science.abo3324,Wang2023MetasurfacePTC}. Together, these effects raise the possibility of jointly controlling the spectrum, emission direction, and intensity of scattered light within a single SPC. Extending this approach to relativistic laser intensities motivates the exploration of plasma-based spatiotemporal structures.

A relativistically driven plasma surface provides a way to create such a spatiotemporal modulation. When an intense laser pulse irradiates a solid target, the plasma formed at its surface can generate high-order harmonics in the reflected light, a process known as high-order harmonic generation (HHG) \cite{PhysRevLett.76.50,Timmis2026EfficiencyOptimized}. In the relativistic oscillating mirror (ROM) model, the reflecting plasma surface oscillates along the target normal at relativistic velocities. Its motion modulates the phase of the reflected field and can generate high-order harmonics \cite{10.1063/1.870766,Lichters_1996,Dromey2006RelativisticLimit,PhysRevE.74.046404}. From a spatiotemporal perspective, this process can be understood as light interacting with a dynamically modulated plasma boundary.

In a single-beam configuration, the same laser pulse both drives the plasma motion and undergoes modulation upon reflection, making it difficult to vary the driving conditions and the properties of the modulated light independently. A pump--probe configuration provides separate control over these two roles \cite{Xia2023AngularDispersed,Wang2025IsolatedAttosecond,Zhang2026OPGPRR}. From the perspective of laser--plasma interactions, a challenge in multibeam configurations is to connect the driven plasma response to both the emission directions of the harmonics and the energy carried by each channel. The SPC framework offers a common physical picture for these processes by relating the optical response to the spatiotemporal periodicity and band structure of the driven plasma \cite{Sharabi:22,PhysRevE.75.046607}.

\begin{figure*}[t]
    \centering
    \includegraphics[width=0.646\textwidth,trim={17.02bp 21.60bp 28.80bp 6.76bp},clip]{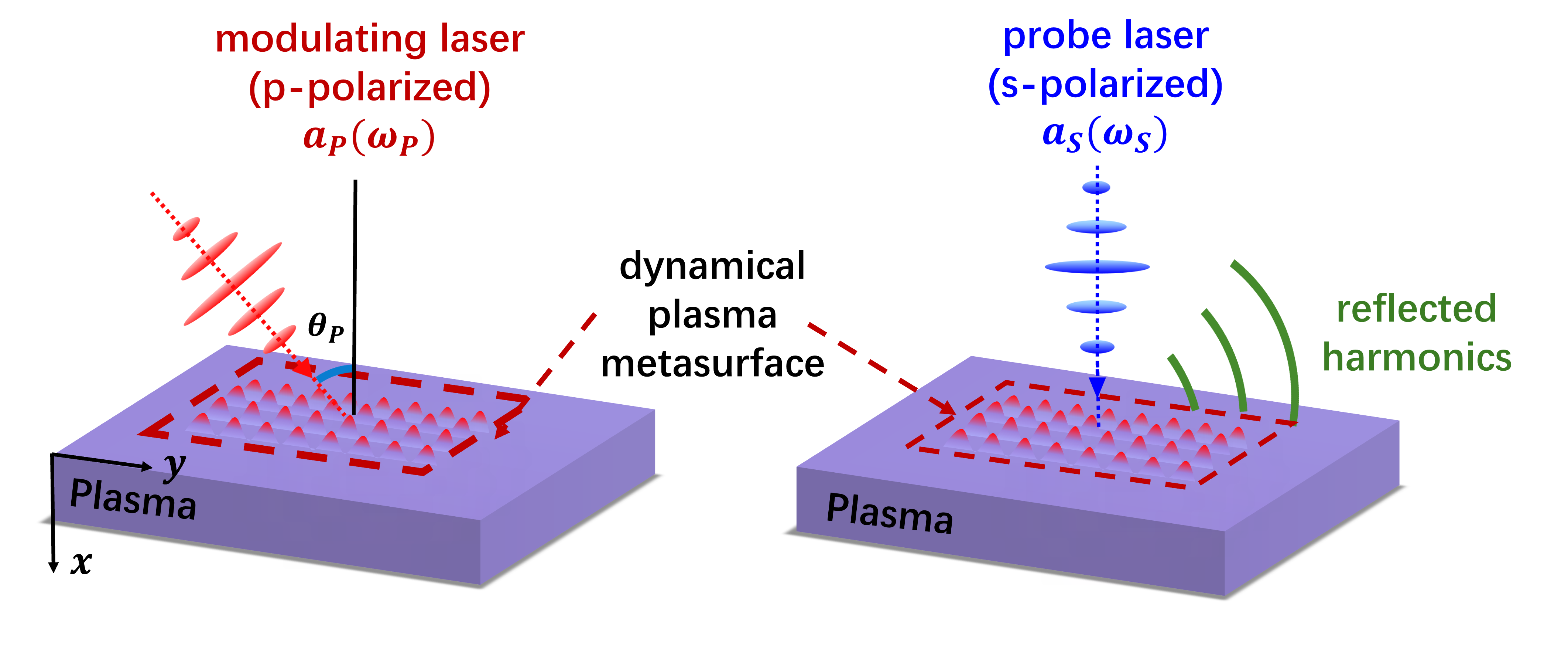}
    \caption{Schematic view of the HHG process from a laser-driven SPPC. An obliquely incident p-polarized modulating laser pulse drives the oscillatory motion of the surface plasma on a solid target, forming a dynamical metasurface (left). A normally incident s-polarized probe pulse interacts with this metasurface and generates reflected high-order harmonics (right).}
    \label{fig:Physicalmodel}
\end{figure*}

In this work, we adopt the SPC framework to describe optical modulation at a relativistically driven plasma surface (Fig.~\ref{fig:Physicalmodel}). We consider an obliquely incident $p$-polarized pump pulse that drives periodic surface motion, creating a spatiotemporal plasma photonic crystal (SPPC), while an $s$-polarized probe interacts with this driven surface and generates high-order harmonics upon reflection. The orthogonal polarizations distinguish the probe-induced harmonics from the emission generated directly by the pump \cite{Lichters_1996}. We show that the reciprocal lattice of the SPPC determines the allowed harmonic frequencies and emission angles. We further find that the probe can draw energy from the pump-driven modulation, a process explained by the complex Floquet bands. The predicted probe energy gain and harmonic spectrum agree well with particle-in-cell (PIC) simulations. This picture also predicts wavelength-dependent probe amplification as the probe accesses amplifying modes of the SPPC. We then extend the SPPC interpretation to the case of two intersecting $p$-polarized pumps.

\section{Physical Model and PIC Simulation Setup}
\label{sec:PIC}
Figure~\ref{fig:Physicalmodel} illustrates the pump--probe interaction at the plasma surface. The $x$ axis is normal to the target, and the $y$ axis lies along its surface in the plane of incidence. In the single-pump simulations, the $p$-polarized pump has wavelength $\lambda_P=800~\mathrm{nm}$ and is incident at $\theta_P=\pi/4$ from the target normal. The normally incident $s$-polarized probe is polarized along $z$ and has wavelength $\lambda_S=347~\mathrm{nm}$ ($\omega_S\simeq2.305\omega_P$), giving a noninteger frequency ratio that helps distinguish the probe sidebands. The normalized peak amplitudes are $a_P=2$ and $a_S=0.5$, with $a_j=eE_j/(m_ec\omega_j)$ for $j=P,S$. The $p$-polarized pump drives the plasma surface, and scattering of the $s$-polarized probe from the resulting modulation generates reflected $s$-polarized harmonics.

To expose how the pump imprints a spatiotemporal modulation on the plasma, we first write its optical phase explicitly. Neglecting the pulse envelope for this qualitative discussion, the normalized vector potential is
\begin{equation}
    \boldsymbol{a}_P
    =
    a_P\sin\Phi_P
    (\hat{y}\cos\theta_P-\hat{x}\sin\theta_P),
    \label{eq:pump_fields}
\end{equation}
where
\begin{equation}
    \Phi_P
    =
    \omega_Pt-k_Px\cos\theta_P-k_Py\sin\theta_P .
    \label{eq:pump_phases}
\end{equation}
Here $\Phi_P$ is the propagation phase of the pump. Consistent with the amplitude definition above, the corresponding electric field follows from $\boldsymbol{E}_P=-(m_ec/e)\partial_t\boldsymbol{a}_P$ and has an amplitude $E_P=a_P\omega_Pm_ec/e$. The pump therefore has a finite surface-normal component, with amplitude $|E_{P,x}|=E_P\sin\theta_P$, which directly drives the surface electrons and imprints its phase on the plasma response.

The pump-driven electron response is governed by the relativistic cold-fluid equations
\begin{align}
    \partial_t n_e+
    \boldsymbol{\nabla}\boldsymbol{\cdot}
    (n_e\boldsymbol{v}_e)&=0,\notag\\
    \left(
    \partial_t+\boldsymbol{v}_e\boldsymbol{\cdot}
    \boldsymbol{\nabla}
    \right)
    (\gamma_em_e\boldsymbol{v}_e)
    &=-e\left(
    \boldsymbol{E}_P+\boldsymbol{E}_{\mathrm{es}}
    +\boldsymbol{v}_e\times\boldsymbol{B}_P
    \right),\notag\\
    \boldsymbol{\nabla}\boldsymbol{\cdot}\boldsymbol{E}_{\mathrm{es}}
    &=-\frac{e(n_e-n_0)}{\varepsilon_0},
    \label{eq:surface_fluid}
\end{align}
where $\boldsymbol{E}_{\mathrm{es}}$ is the electrostatic restoring field, $\gamma_e=[1-(v_{e,x}^2+v_{e,y}^2)/c^2]^{-1/2}$ is the electron Lorentz factor, and $\boldsymbol{\nabla}=\hat{x}\partial_x+\hat{y}\partial_y$ in the present two-dimensional geometry.

To illustrate analytically how the pump phase is imprinted on the surface-density response, we take the leading-order small-excursion limit of Eq.~(\ref{eq:surface_fluid}). We write $n_e=n_0(x)+\delta n_e$, set $\gamma_e\simeq1$, and retain terms linear in the pump field. The continuity and momentum equations then reduce to
\begin{align}
    \partial_t\delta n_e
    +\boldsymbol{\nabla}\boldsymbol{\cdot}
    (n_0\boldsymbol{v}_e)&=0,\notag\\
    m_e\partial_t\boldsymbol{v}_e
    &=-e(\boldsymbol{E}_P+\boldsymbol{E}_{\mathrm{es}}).
    \label{eq:linearized_surface_fluid}
\end{align}
Within a local surface-layer approximation, in which the electrostatic restoring response is represented by the local plasma frequency, these equations give
\begin{equation}
    \left(\partial_t^2+\omega_{pe}^2\right)\delta n_e
    =
    \frac{e}{m_e}
    \boldsymbol{\nabla}\boldsymbol{\cdot}
    (n_0\boldsymbol{E}_P).
    \label{eq:driven_density}
\end{equation}
Here $\omega_{pe}=\sqrt{n_0e^2/(\varepsilon_0m_e)}$ is the local plasma frequency associated with the equilibrium density. The source term can be separated as
\begin{align}
    \boldsymbol{\nabla}\boldsymbol{\cdot}
    (n_0\boldsymbol{E}_P)
    &=
    \boldsymbol{E}_P\boldsymbol{\cdot}
    \boldsymbol{\nabla}n_0
    +n_0\boldsymbol{\nabla}\boldsymbol{\cdot}
    \boldsymbol{E}_P.
\end{align}
The field-divergence term vanishes for the prescribed transverse plane-wave pump. With $n_0=n_0(x)$, the source term is then approximated by
\begin{equation}
    \boldsymbol{\nabla}\boldsymbol{\cdot}
    (n_0\boldsymbol{E}_P)
    \simeq
    E_{P,x}\frac{\mathrm{d}n_0}{\mathrm{d}x}.
    \label{eq:surface_density_source}
\end{equation}

At a fixed depth away from resonance, Eq.~(\ref{eq:driven_density}) has the local particular solution
\begin{equation}
    \delta n_e
    \simeq
    \frac{e}{m_e}
    \frac{\mathrm{d}n_0}{\mathrm{d}x}
    \frac{E_P\sin\theta_P}
    {\omega_{pe}^2-\omega_P^2}
    \cos\Phi_P .
    \label{eq:density_solution}
\end{equation}
Equation~(\ref{eq:density_solution}) shows that, at oblique incidence, the surface-density perturbation inherits the pump phase, giving rise to a spatiotemporal modulation of the effective permittivity. This leading-order response is driven by the surface-normal electric field and vanishes at normal incidence, where higher-order ponderomotive forces can still drive periodic density variations. In the relativistic regime, the motion of the electron layer across the steep surface density gradient is strongly nonlinear. This nonlinear motion reshapes the evolution of the density and permittivity within each optical cycle. The overall modulation pattern approximately repeats from one pump cycle to the next. The linearized solution illustrates the origin of this recurrence. We obtain the nonlinear waveform used to reconstruct the permittivity from pump-only PIC simulations.

We use the two-dimensional PIC code EPOCH \cite{arber2015contemporary} to simulate an overdense target with a truncated exponential preplasma. Section~1 of Supplement~1 gives the pulse, target, and numerical parameters. The pump-only run provides the driven plasma response. The probe-only run provides the baseline for the pump-induced gain. The pump-plus-probe run tests the predicted gain and spectrum independently. The incident probe in the SPPC calculation is specified by its injection parameters.

Using the pump-only PIC density $n_e$ and cell-averaged kinetic energy $\langle K\rangle$, we obtain the scalar response for the $s$-polarized probe,
\begin{align}
    \bar{\gamma}(x,y,t)
    &=1+\frac{\langle K(x,y,t)\rangle}{m_ec^2},
    \label{eq:effective_gamma}\\
    \varepsilon_{\mathrm{eff}}(x,y,t;\omega)
    &=1-\frac{e^2n_e(x,y,t)}
    {\varepsilon_0m_e\bar{\gamma}(x,y,t)\omega^2}.
    \label{eq:effective_permittivity}
\end{align}

\begin{figure}[h]
    \centering
    \includegraphics[width=1\linewidth]{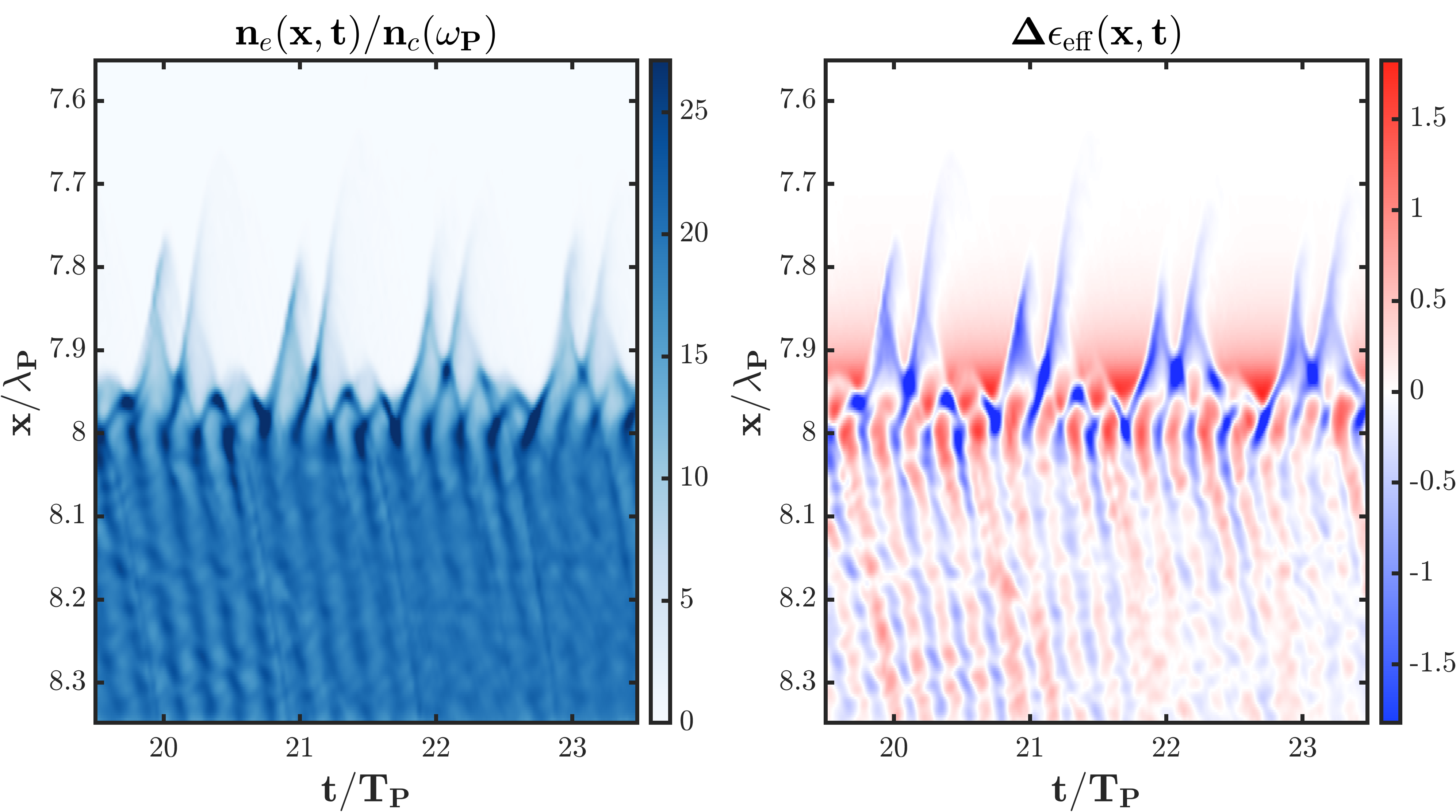}
    \caption{Spatiotemporal modulation of the plasma surface driven by a single p-polarized laser, extracted from the PIC simulation at a fixed transverse position $y=y_0$. The left panel shows the electron density $n_e(x,t)$ normalized to the critical density $n_c(\omega_P)$. The oscillating plasma boundary and the density structures below the surface approximately recur from one pump cycle to the next in the $x$-$t$ plane. The right panel shows the effective permittivity modulation $\Delta\varepsilon_{\mathrm{eff}}(x,t)$ evaluated at the probe frequency $\omega_S$.}
    \label{fig:permittivity_modulation}
\end{figure}

Figure~\ref{fig:permittivity_modulation} shows the density and permittivity modulation at a fixed $y$ position ($y=y_0$). The right panel shows $\Delta\varepsilon_{\mathrm{eff}}(x,t)$ obtained from $\varepsilon_{\mathrm{eff}}(x,y_0,t;\omega_S)$ by subtracting its time average over the four pump periods shown.

\section{Harmonic Generation from Spatiotemporal Photonic Crystals}
The cycle-to-cycle recurrence seen in Fig.~\ref{fig:permittivity_modulation}, together with the tangential periodicity imposed by the pump phase in Eq.~(\ref{eq:pump_phases}), makes the driven plasma surface an SPPC. We use this description to derive the selection rules for harmonic frequencies and emission angles and to calculate the probe energy gain and harmonic spectrum.

\subsection{Field Transformation under a Geometric Time-Space Interface Rotation}
\begin{figure}[h]
    \centering
    \includegraphics[width=0.285\linewidth,trim={26.40bp 0.22bp 17.02bp 0.00bp},clip]{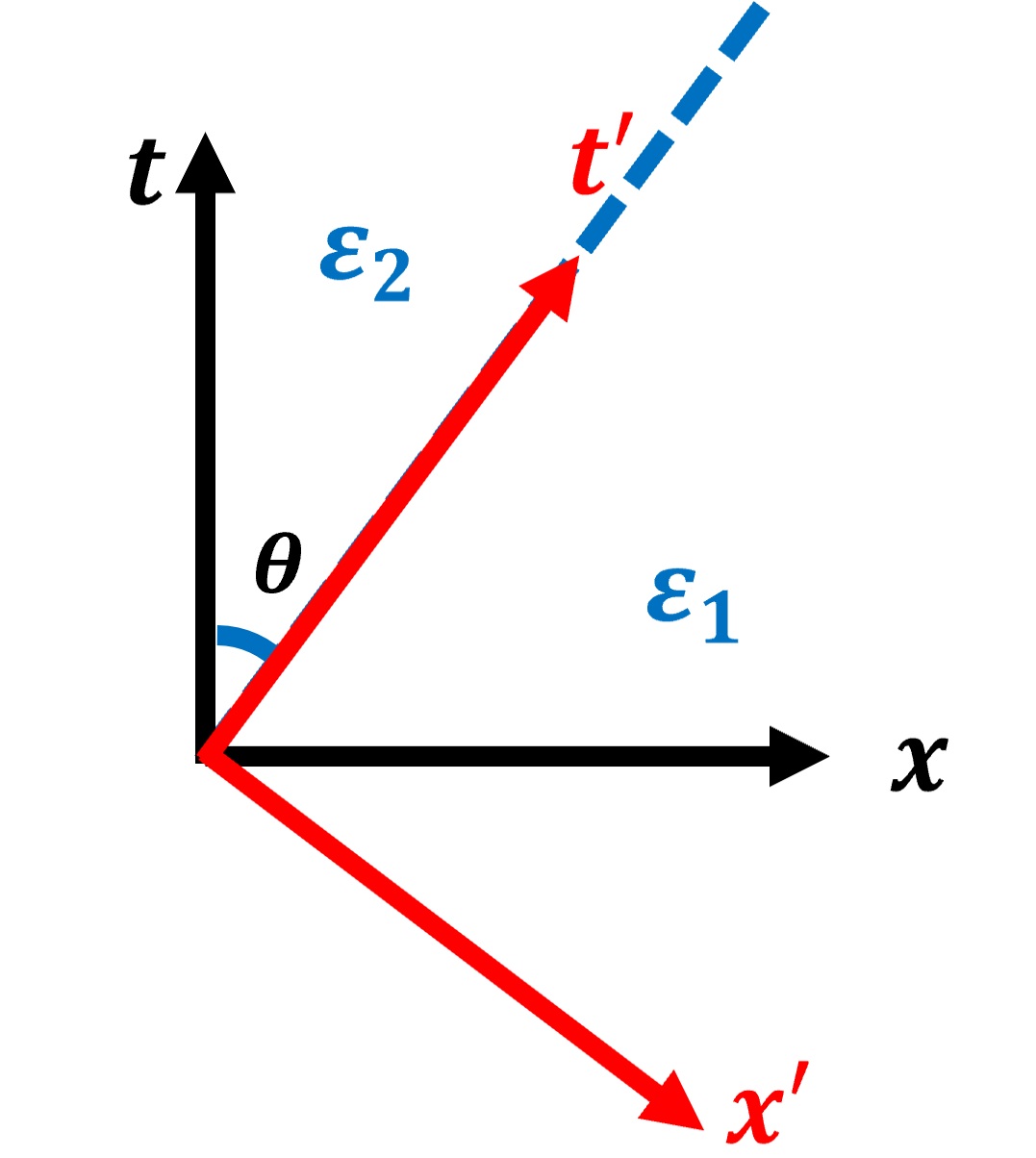}
    \caption{A time-space interface describes a local change in permittivity in the $t$-$x$ plane. Under an appropriate geometric time-space transformation, the new time axis ($t'$) becomes parallel to the interface, thereby transforming the time-space interface into a purely spatial interface.}
    \label{fig:xt}
\end{figure}

The periodic permittivity modulation of an SPC can be modeled as a collection of local time-space interfaces. We therefore begin by considering the interaction of the probe with a single interface (Fig.~\ref{fig:xt}). For simplicity, we restrict the spatial variation to the $x$ direction. The time- and space-dependent permittivity $\varepsilon(t,x)$ can then be represented by a discontinuous profile across an interface in the $x-t$ plane. The media on the two sides of the interface have different permittivities, and the transition can be written with a Heaviside function,
\begin{equation}\label{epsiloninterface}
    \varepsilon(t,x)=\varepsilon_1+(\varepsilon_2-\varepsilon_1)\Theta(\beta t-x),
\end{equation}
where $\beta\equiv\tan\theta$ is the velocity of the interface in the $t-x$ coordinates. Here we set $c=1$ for convenience. In general, $\beta$ may depend on both time and space. However, within a sufficiently short time interval, the interface motion can always be approximated as locally uniform.

A geometric rotation, rather than a Lorentz transformation, can be introduced to transform the time-space interface into a purely spatial interface,
\begin{equation}
    x'^\mu=\Lambda^\mu_{\nu}x^\nu,
\end{equation}
where $x^\mu=(t,x,y,z)$. The rotation matrix is
\begin{equation}\label{xxtrans}
\Lambda=
    \begin{pmatrix}
    \cos\theta & \sin\theta &0&0\\
    -\sin\theta & \cos\theta &0&0\\
    0&0&1&0\\
    0&0&0&1
    \end{pmatrix}.
\end{equation}
The coordinate derivatives transform as
\begin{equation}\label{coordinate_derivatives}
    \frac{\partial}{\partial x'^\mu}=(\Lambda^{-1})^{\nu}_{\mu}\frac{\partial}{\partial x^{\nu}},
\end{equation}
where $\Lambda^{-1}$ is the inverse of $\Lambda$.

After this rotation, the permittivity becomes independent of the new time coordinate $t'$,
\begin{equation}
     \varepsilon(t,x)=\varepsilon'(x').
\end{equation}
Therefore, translation invariance is restored along the $t'$ direction, as shown in Fig.~\ref{fig:xt}. From Eq.~(\ref{xxtrans}), the generalized frequency and wave vector in the rotated coordinates are
\begin{equation}\label{generalizedfrequency}
    k'^\mu=(\Lambda^{-1})^\mu_{\nu}k^\nu,
\end{equation}
where $k^\mu=(\omega,k_x,k_y,k_z)$. Biancalana \textit{et al.} have shown that the generalized frequency $\omega'$ is conserved during scattering from a time-space interface \cite{PhysRevE.75.046607}. This conservation law can be understood from the restored translation symmetry along the $t'$ direction: $\omega'$ is the generator associated with this symmetry.

The electromagnetic field tensor transforms under the same rotation,
\begin{equation}\label{em_field}
    F'_{\mu\nu}=(\Lambda^{-1})^{\alpha}{}_{\mu}
    (\Lambda^{-1})^{\beta}{}_{\nu}F_{\alpha\beta},
\end{equation}
where $F_{\mu\nu}=\partial_\mu A_\nu-\partial_\nu A_\mu$ and $A_\mu$ is the electromagnetic four-potential.

\subsection{Frequencies of Light Emitted from the SPPC}
\label{sec:single_pump_frequencies}
We next derive the frequency components generated by a periodically driven time-space interface. {For convenience, we consider the scattering of a monochromatic s-polarized probe from a prescribed modulation background generated by the stronger p-polarized pump. The probe is a plane wave propagating along the $x$ direction, with its electric field polarized along $z$.} This description is valid locally during a short interval in which the interface velocity can be approximated by $\beta=\tan\theta$.

Within the effective-medium description, Maxwell's equations for the probe field are
\begin{align}\label{maxwell}
\nabla\times E+\frac{\partial B}{\partial t}&=0,\\
\nabla\times H-\frac{\partial D}{\partial t}&=0,
\end{align}
{For the frequency-selection derivation, we neglect the weak probe feedback on the pump-driven plasma and evaluate the absolute permittivity $\varepsilon(t,x)=\varepsilon_0\varepsilon_{\mathrm{eff}}(x,y_0,t;\omega_S)$ at a fixed transverse position $y_0$. The instantaneous constitutive relations are}
\begin{align}
   D&=\varepsilon(t,x)E,\label{eq:instantaneous_constitutive}\\
   H&=B/\mu_0.
\end{align}
In the geometrically rotated coordinate system, the permittivity depends only on $x'$. Because the pump-driven plasma response approximately recurs from one optical cycle to the next, $\varepsilon'(x')$ can be treated as periodic over the relevant interaction interval. We denote the period in the rotated coordinate by $l'$,
\begin{equation}
    \varepsilon'(x')=\varepsilon'(x'+l').
\end{equation}
For a modulation frequency \(\omega_P\), the temporal period in the original coordinates is \(T_P=2\pi/\omega_P\). Under the rotation \(x'=-t\sin\theta+x\cos\theta\), a temporal translation \(\Delta t=T_P\) at fixed \(x\) corresponds to \(\Delta x'=-T_P\sin\theta\). Since the permittivity is independent of \(t'\), its period along \(x'\) is therefore \(l'=|\Delta x'|=2\pi|\sin\theta|/\omega_P\).

{Expressing the permittivity in the rotated coordinates, we write the constitutive relation as}
\begin{equation}\label{eq:rotated_constitutive}
    {D=\varepsilon'(x')E.}
\end{equation}

Applying the transformations in Eqs.~(\ref{coordinate_derivatives}) and (\ref{em_field}) and taking $E_z'(x',t')=E_{z0}'(x')e^{-i\omega't'}$ and $B_y'(x',t')=B_{y0}'(x')e^{-i\omega't'}$ gives the following wave equation for the electric-field amplitude (see Section~2 of Supplement~1).
\begin{align}
    &\frac{\mathrm{d}^2E_{z0}'}{\mathrm{d}x'^2}
    -\frac{\mu_0\sin^2\theta\,\partial_{x'}\varepsilon'
    +2i\omega'\sin\theta\cos\theta(1+\mu_0\varepsilon')}
    {\cos^2\theta-\mu_0\varepsilon'\sin^2\theta}
    \frac{\mathrm{d}E'_{z0}}{\mathrm{d}x'}\notag\\
    &+\frac{(\mu_0\varepsilon'\cos^2\theta-\sin^2\theta)\omega'^2
    -i\omega'\mu_0\sin\theta\cos\theta\,\partial_{x'}\varepsilon'}
    {\cos^2\theta-\mu_0\varepsilon'\sin^2\theta}E'_{z0}=0.\label{helmhotzeq}
\end{align}
Since the coefficients in Eq.~(\ref{helmhotzeq}) are periodic in $x'$ with period $l'$, Floquet's theorem states that the field can be expanded in Bloch modes. For a general wave packet one may write
\begin{equation}
    E_{z0}'(x')=\int_{\mathrm{BZ}} \mathrm{d}q\, A(q)\,u_q(x')e^{iqx'},
\end{equation}
where $q$ is the Bloch wavenumber, $\mathrm{BZ}$ denotes the first Brillouin zone
of the reciprocal lattice defined by $G=2\pi/l'$, and
$u_q(x')=u_q(x'+l')$ is periodic. Equivalently,
\begin{equation}
    u_q(x')=\sum_n e_{q,n}'e^{iG_nx'},\qquad G_n=\frac{2\pi n}{l'} .
\end{equation}
Thus each Bloch component contains Fourier harmonics with wavenumbers
$q+G_n$.    

For the following frequency-selection argument, it is sufficient to consider
one incident Bloch component, because different Bloch components evolve
independently under the linear Floquet problem. The Bloch wavenumber of this
component is fixed by the incident probe laser. Specifically, once the frequency
and propagation direction of the incident probe laser are specified in the original
coordinates, its generalized frequency and wavenumber in the rotated
coordinates are determined by the rotation. For an incident probe laser component
with frequency $\omega_S$ and wavenumber $k_S$ along the $x$ direction, one has
\begin{equation}
    \omega'_S=\omega_S\cos\theta-k_S\sin\theta,\qquad
    k'_S=\omega_S\sin\theta+k_S\cos\theta .\label{incidentwave}
\end{equation}
Thus, for a fixed incident branch, $\omega_S$ and $k'_S$ are in one-to-one
correspondence.

For this incident Bloch component, the periodic part can be expanded in
reciprocal-lattice harmonics,
\begin{equation}
    E_{z0}'(x')=\sum_n e_n' e^{i(k'_S+G_n)x'},
    \qquad G_n=\frac{2\pi n}{l'} .
\end{equation}
Similarly,
\begin{equation}
    B_{y0}'(x')=\sum_n b_n' e^{i(k'_S+G_n)x'} .
\end{equation}

Transforming back to the original coordinates and using Eq.~(\ref{incidentwave}), we obtain
\begin{align}
    E_z&=\cos\theta E_{z0}'e^{-i\omega_S't'}-\sin\theta B_{y0}'e^{-i\omega_S't'} \notag\\
    &=\sum_n(e_n'\cos\theta-b_n'\sin\theta)
    e^{i[(k_S+G_n\cos\theta)x-(\omega_S+G_n\sin\theta)t]}.
\end{align}
The frequency of the $n$-th component is therefore
\begin{equation}
    \omega_n=\omega_S+G_n\sin\theta
    =\omega_S+\frac{2\pi n}{l'}\sin\theta .
\end{equation}
For the frequency-selection derivation, we take $\sin\theta>0$ without loss of generality; reversing the interface orientation is equivalent to relabeling $n\rightarrow-n$. Thus, $l'=2\pi\sin\theta/\omega_P$, and the generated frequencies are
\begin{equation}
    \omega_n=\omega_S+n\omega_P .
\end{equation}
Therefore, a periodically driven time-space interface converts the incident probe laser into {harmonics} separated by the modulation frequency. Although the derivation above assumes a local interface with constant velocity, a more complicated plasma motion can be decomposed into sufficiently short intervals, with the generated field from one interval serving as the incident field for the next.

{This derivation establishes the frequency-selection rule from the modulation periodicity, without determining the frequency-dependent response or intensity of individual harmonics. The generated components have their own frequency-dependent plasma responses and may undergo further frequency conversion. Each conversion shifts the frequency by an integer multiple of $\omega_P$, so the allowed frequencies remain $\omega_S+n\omega_P$. The quantitative harmonic spectrum is calculated below using a frequency-dependent response model, as detailed in Section~3 of Supplement~1.}

\subsection{Emission Angles of Harmonics}
\label{sec:single_pump_angles}
\begin{figure}[h]
    \centering
    \includegraphics[width=0.829\linewidth,trim={2.40bp 10.04bp 10.69bp 5.45bp},clip]{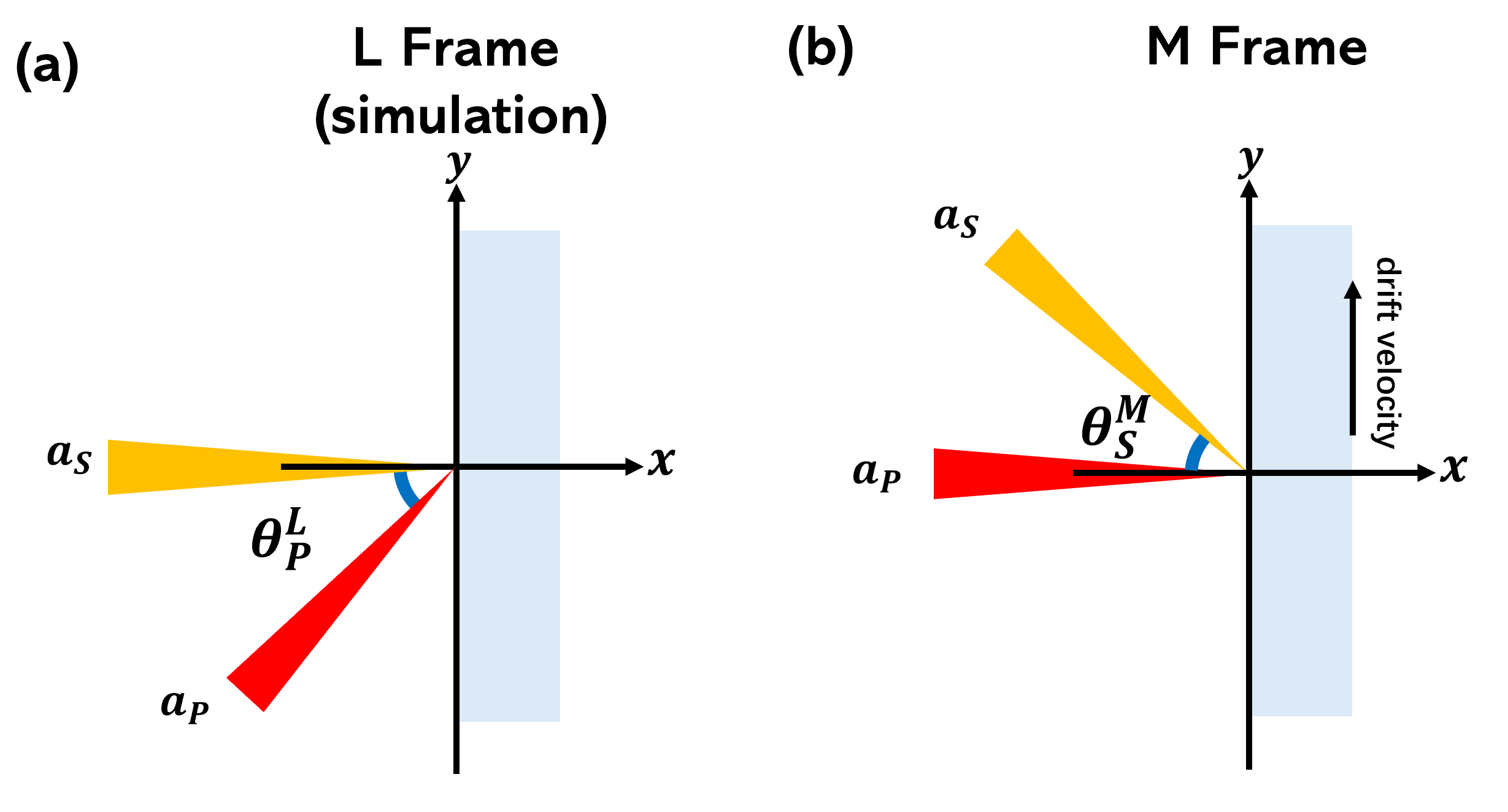}
    \caption{Relative orientations between the laser pulses and the target in the laboratory frame $L$ \textbf{(a)} and the moving frame $M$ \textbf{(b)}. The red and yellow beams represent the p- and s-polarized lasers, respectively. With signed incidence angles, the s-polarized laser satisfies $\theta_S^M=-\theta_P^L$.}
    \label{LP}
\end{figure}

We then obtain the harmonic emission angles for a plasma surface driven by a single obliquely incident p-polarized laser using conservation of the generalized frequency. In the laboratory frame $L$, the p-polarized driver is incident at an angle $\theta_P^L$, whereas the s-polarized probe is normally incident. To analyze the conserved quantities more clearly, we introduce a Lorentz transformation to a moving frame $M$, in which the p-polarized driver becomes normally incident, as shown in Fig.~\ref{LP}. The drift velocity is $v_d=\sin\theta_P^L$. Within the plane-wave and locally planar-surface approximation, the system in the $M$ frame is translationally invariant along the target surface, and the pump-driven plasma response has no $y$ dependence.

Under this translational-invariance approximation, the interface is temporal in the $y$ direction. It can be written as
\begin{equation}
    \varepsilon=\varepsilon_1+(\varepsilon_2-\varepsilon_1)\Theta(\beta_x t-x)\Theta(\beta_y t-y),
    \;\beta_y\to+\infty.
\end{equation}
In the limit $\beta_y\to+\infty$, the second step function reduces to $\Theta(t)$. A geometric rotation in the $t$-$y$ plane then gives
\begin{equation}
    \begin{pmatrix}
    t'\\y'
    \end{pmatrix}
    =
    \begin{pmatrix}
    0&1\\
    -1&0
    \end{pmatrix}
    \begin{pmatrix}
    t\\y
    \end{pmatrix}.
\end{equation}
Correspondingly,
\begin{equation}
    \begin{pmatrix}
    \omega'\\k_y'
    \end{pmatrix}
    =
    \begin{pmatrix}
    0&-1\\
    1&0
    \end{pmatrix}
    \begin{pmatrix}
    \omega\\k_y
    \end{pmatrix}.
\end{equation}
Thus, conservation of the generalized frequency $\omega'$ in the $M$ frame is equivalent to conservation of $-k_y$, where $k_y$ is the $y$-direction wavenumber in the $M$ frame.

Transforming this conserved quantity back to the laboratory frame gives
\begin{equation}\label{selectionrules_in_S}
   k_{n,y}^L=\frac{k_{n,y}^M+v_d\omega_n^M}{\sqrt{1-v_d^2}}
   =n k_P^L\sin\theta_P^L.
\end{equation}
Here $n$ labels the pump-modulated harmonics of the probe, and the superscripts $L$ and $M$ denote the laboratory and moving frames, respectively. Combining this transverse-momentum selection rule with the frequency-selection rule and the vacuum dispersion relation gives the harmonic positions and emission angles,
\begin{align}
   & (k_{n,x}^L,k_{n,y}^L)=\omega_n(\cos\theta_n,\sin\theta_n),\notag\\
   & \omega_n=\omega_S+n\omega_P,\notag\\
   & \sin\theta_n=\frac{k_{n,y}^L}{k_n^L}=\frac{n\omega_P\sin\theta_P}{\omega_S+n\omega_P}.
   \label{position_single}
\end{align}
The dependence of $\theta_n$ on the harmonic frequency shows that the single-pump harmonic emission is angularly dispersed.

\begin{figure}[htbp]
    \centering
      \includegraphics[width=0.88\linewidth]{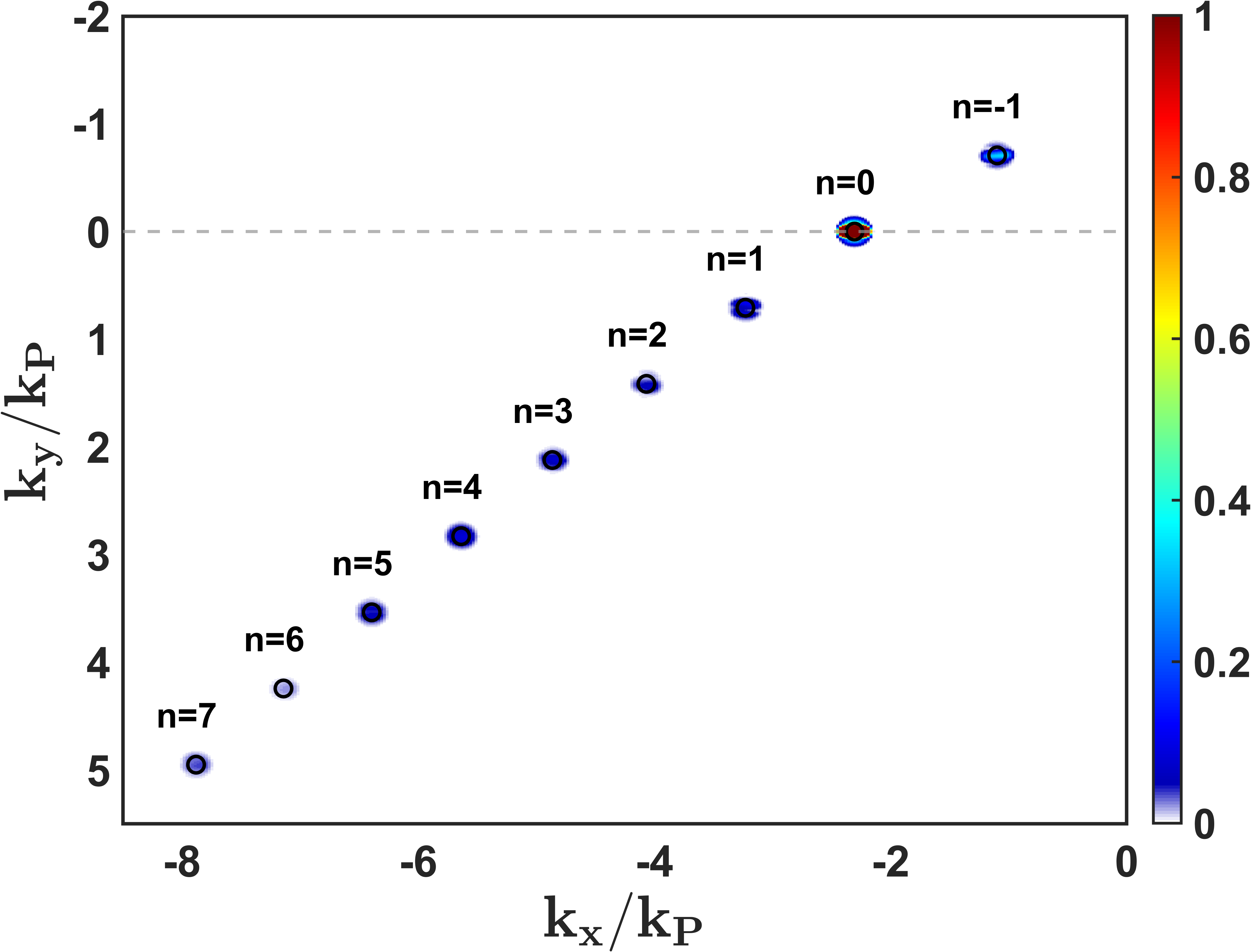}
    \caption{Comparison of the harmonic positions obtained from the single-pump PIC simulation and the theoretical prediction. The background shows $|E_z(k_x,k_y)|$ in $k$ space, and the circles mark the harmonic positions predicted by Eq.~(\ref{position_single}). The integer $n$ denotes the harmonic order associated with the SPPC driven by the p-polarized pump. Since the reflected harmonic components are analyzed, $k_x$ is negative.}
    \label{harmonics_SPC_single}
\end{figure}

Figure~\ref{harmonics_SPC_single} compares the harmonic positions predicted by Eq.~(\ref{position_single}) with the PIC simulation for the single-pump configuration described in Sec.~\ref{sec:PIC}. The background shows $|E_z(k_x,k_y)|$ in $k$ space obtained from the reflected field in the PIC simulation, while the circles mark the theoretical positions.

\subsection{Emission Efficiency Enhancement}
In a photonic time crystal (PTC), temporal modulation can open momentum gaps with complex Floquet eigenfrequencies \cite{Lustig:18}. Within these gaps, growing modes draw energy from the modulation \cite{doi:10.1126/science.abo3324,Sharabi:22}. We use this amplification mechanism as a starting point to examine how the pump-driven SPPC amplifies the probe and enhances its harmonic emission.

\begin{figure}[h]
    \centering
    \includegraphics[width=0.800\linewidth]{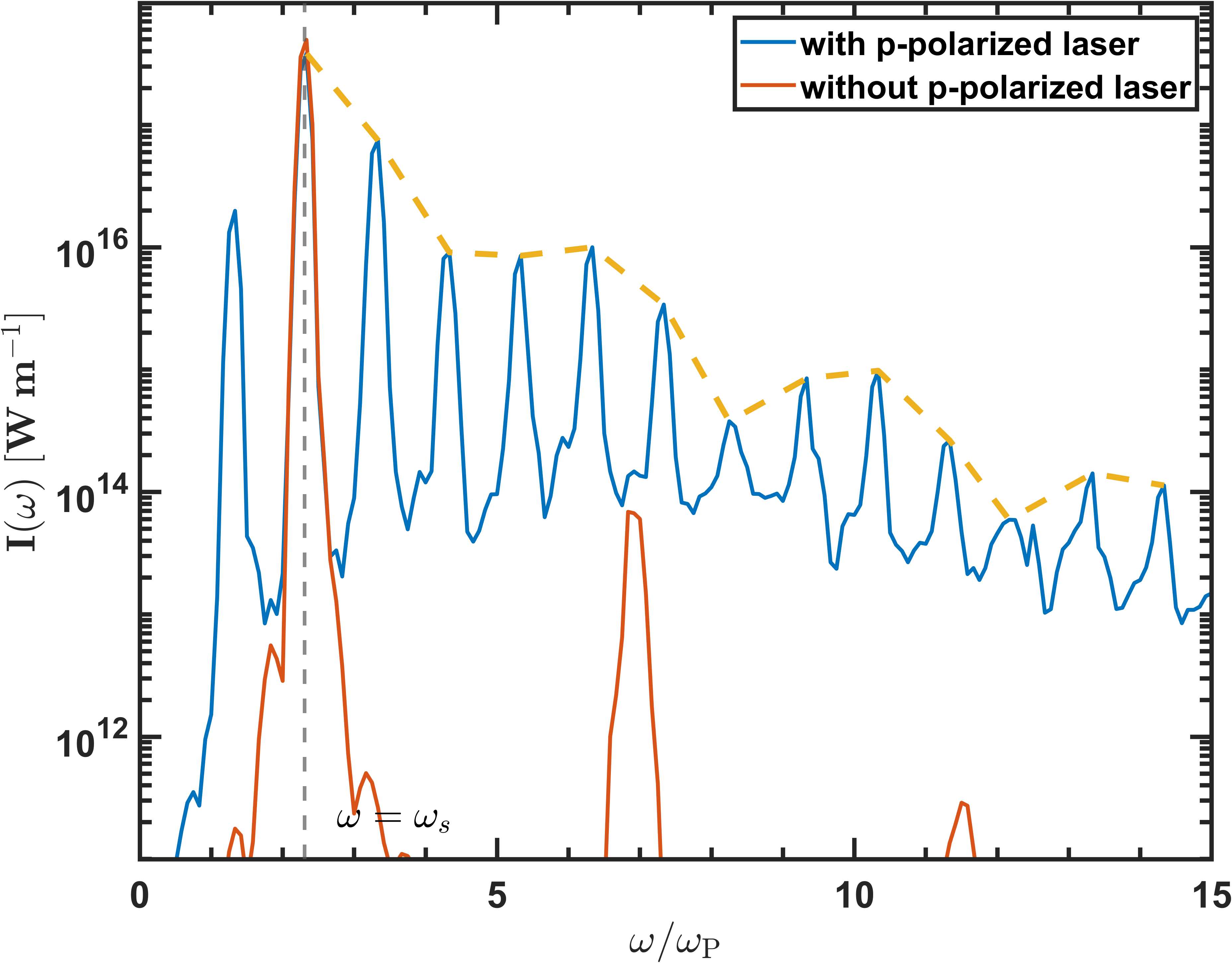}
    \caption{{Angle-integrated harmonic spectra of the s-polarized probe with and without the p-polarized pump.} The blue line shows the spectral intensity with the p-polarized driver present, and the orange line shows the harmonics generated by the s-polarized {probe} alone.}
    \label{Iw}
\end{figure}

Figure~\ref{Iw} compares the angle-integrated spectra from the pump-plus-probe and probe-only PIC simulations (blue and orange, respectively). With the pump present, distinct peaks appear at $\omega_S+n\omega_P$, and the high-frequency spectral intensity increases by several orders of magnitude. The yellow dashed line connects the peaks and traces the nonmonotonic spectral envelope. We attribute this modulation to interference between subcycle emission bursts from the compact electron layer formed by preplasma truncation (Section~1 of Supplement~1) \cite{PhysRevE.100.053207,li2022experimental}.

\begin{figure}[h]
    \centering
    \includegraphics[width=0.867\linewidth,trim={29.00bp 35.00bp 20.00bp 55.00bp},clip]{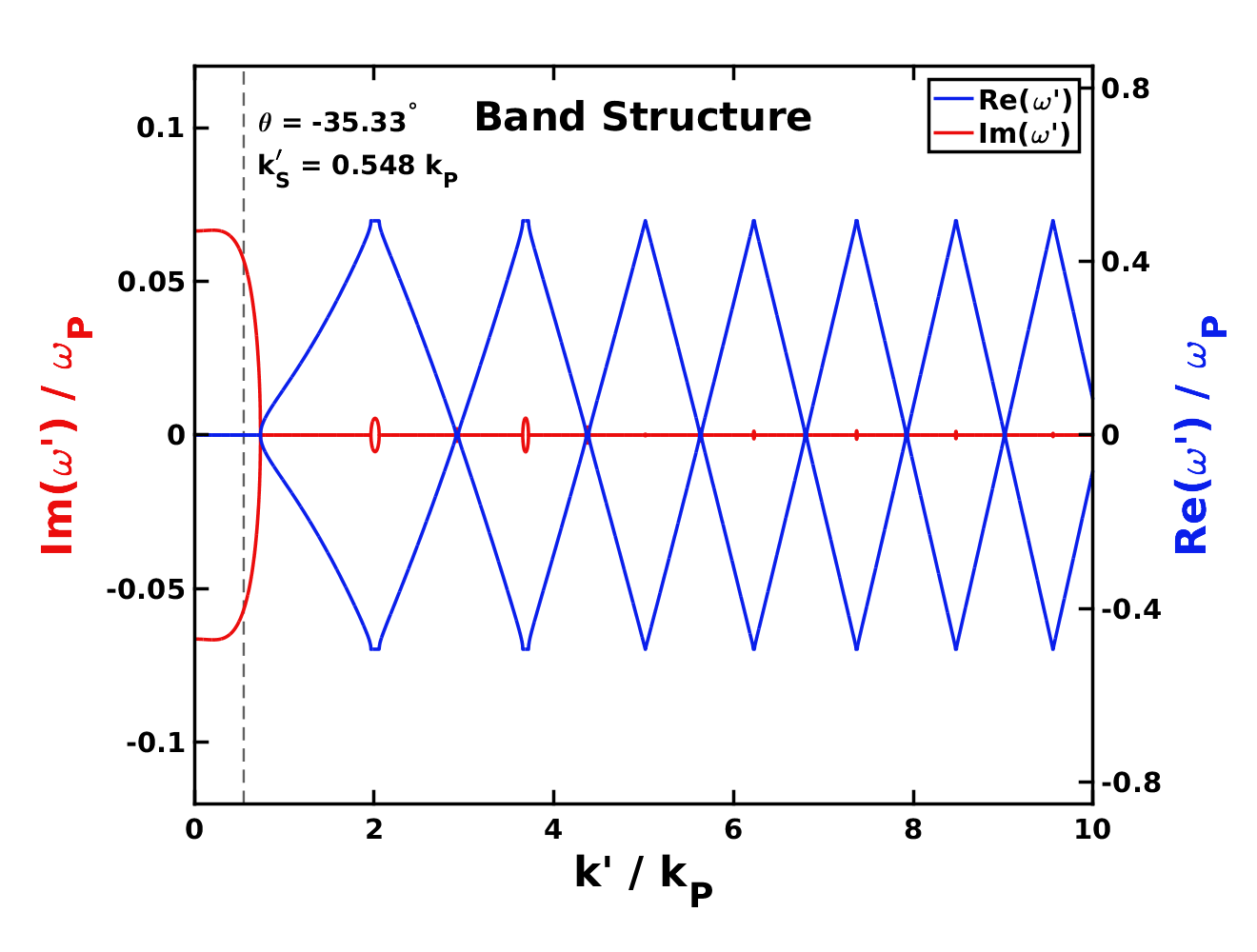}
    \caption{Complex band structure {reconstructed from the pump-only PIC electron dynamics} at {the representative transverse position $y=23.253\lambda_P$ during the second pump period} and expressed in the generalized coordinates $(\omega',k')$. The blue curve shows $\mathrm{Re}(\omega')/\omega_P$ on the right axis, and the red curve shows $\mathrm{Im}(\omega')/\omega_P$ on the left axis. The gray dashed line marks the generalized probe wavenumber {$k_S'=0.548k_P$ obtained for $\theta=-35.33^\circ$.}}
    \label{bandstructure}
\end{figure}

For the representative region shown in Fig.~\ref{bandstructure}, we periodically extend a one-pump-period window of $\varepsilon_{\mathrm{eff}}(x,y,t;\omega)$ from Eq.~(\ref{eq:effective_permittivity}) and calculate the complex Floquet band structure using the transfer-matrix method \cite{PhysRevB.68.045121,Lustig:18}. {We track the temporal evolution of the electron layer in the pump-only PIC simulation and determine its instantaneous velocity $\beta(t)=\mathrm{d}x_{\mathrm{ref}}/(c\,\mathrm{d}t)$ from adjacent time slices. At the selected time, $\beta=-0.709$, giving the rotation angle $\theta=\arctan\beta=-35.33^\circ$.} Using this angle, the complex dispersion is transformed to the generalized coordinates $(\omega',k')$ according to Eq.~(\ref{generalizedfrequency}).

For the incident s-polarized probe, $k_S=2.305k_P$. Using the rotation angle extracted from the critical-surface motion and the $c=1$ convention adopted above, Eq.~(\ref{generalizedfrequency}) gives $k_S'=\omega_S\sin\theta+k_S\cos\theta={0.548k_P}$, as indicated by the gray dashed line in Fig.~\ref{bandstructure}. The incident probe therefore lies inside the first momentum gap, where the two generalized-frequency branches form a complex-conjugate pair. The branch with $\mathrm{Im}(\omega')>0$ grows by drawing energy from the time-dependent plasma modulation sustained by the p-polarized pump. Specifically, for a Floquet mode proportional to $\exp(-i\omega't')$, its field energy grows as $\exp[2\mathrm{Im}(\omega')t']$, so that $2\mathrm{Im}(\omega')$ directly gives the temporal energy-growth rate in the generalized coordinates.

The role of the time-space rotation is important. In a pure PTC, a momentum gap mainly leads to temporal growth. In an SPC, however, the generalized frequency and wave vector are mixtures of the physical frequency and wave vector. As a result, a gap in the generalized band structure corresponds, in the laboratory coordinates, to a combination of temporal amplification and spatial attenuation. For low-frequency components, the temporal growth may therefore be partly offset by spatial decay during propagation. Since the incident probe lies inside the first momentum gap, the p-laser-driven modulation can amplify components near the probe frequency through the imaginary part of the generalized frequency and simultaneously couple this amplified response to sidebands at $\omega_S+n\omega_P$. Because these higher-frequency sidebands lie outside the low-frequency gap, they suffer much weaker attenuation. In this way, energy stored in the driven plasma modulation can be transferred into high-frequency harmonic radiation, consistent with the spectrum shown in Fig.~\ref{Iw}.

To evaluate this amplification quantitatively, we first calculate the total energy transferred from the pump-driven plasma modulation to the s-polarized field using the local complex band structure. {Figure~\ref{bandstructure} shows one representative local band, whereas the gain calculation samples the interaction region at successive points} $(x_i,y_j,t_p)$, where the layers $\{\Delta x_i\}$ cover the probe skin depth, the subregions $\{\Delta y_j\}$ cover the transverse overlap between the probe and the pump-driven plasma modulation, and $\{\Delta T_p\}$ are the selected modulation intervals. {The sampling increments used here are $\Delta x=0.01\lambda_P$, $\Delta y=0.05\lambda_P$, and $\Delta t=0.025T_P$. At each time slice, a window centered on that slice and spanning a total of one pump period is periodically extended to define the local Floquet problem.} At each $(x_i,y_j,t_p)$, the effective permittivity and complex band structure are reconstructed from the local plasma response. {After transforming the Floquet mode back to the laboratory coordinates, the local energy-growth rate $g^{\,\mathrm{net}}_{ijp}$ is determined by $2\operatorname{Im}(\omega)$, while the incident probe energy density $u_{S,\mathrm{in}}(x_i,y_j,t_p)$ is calculated from the prescribed probe field after accounting for the spatial energy-density attenuation governed by $2\operatorname{Im}(k_x)$.}

The local energy gain is evaluated using the probe energy density entering the same spatial and temporal element,
\begin{equation}
    \dot{u}^{\,\mathrm{gain}}_{S,ijp}
    =
    g^{\,\mathrm{net}}_{ijp}
    u_{S,\mathrm{in}}(x_i,y_j,t_p).
\end{equation}
The cumulative predicted energy gain up to the end of the $N$th modulation interval is
\begin{equation}
    \Delta W_{S,\mathrm{pred}}(t_N)
    =
    \sum_{p=1}^{N}
    \sum_{j=1}^{N_y}
    \sum_{i=1}^{N_x}
    \dot{u}^{\,\mathrm{gain}}_{S,ijp}
    \Delta x_i\,\Delta y_j\,\Delta T_p .
    \label{eq:cumulative_energy_gain}
\end{equation}
The field energy remaining within the interaction region after each modulation interval is carried over as part of the input to the next interval, thereby accounting for the persistence and accumulation of field energy across successive pump periods.

\begin{figure}[h]
    \centering
    \includegraphics[width=0.990\columnwidth,trim={2.00bp 27.00bp 13.00bp 1.00bp},clip]{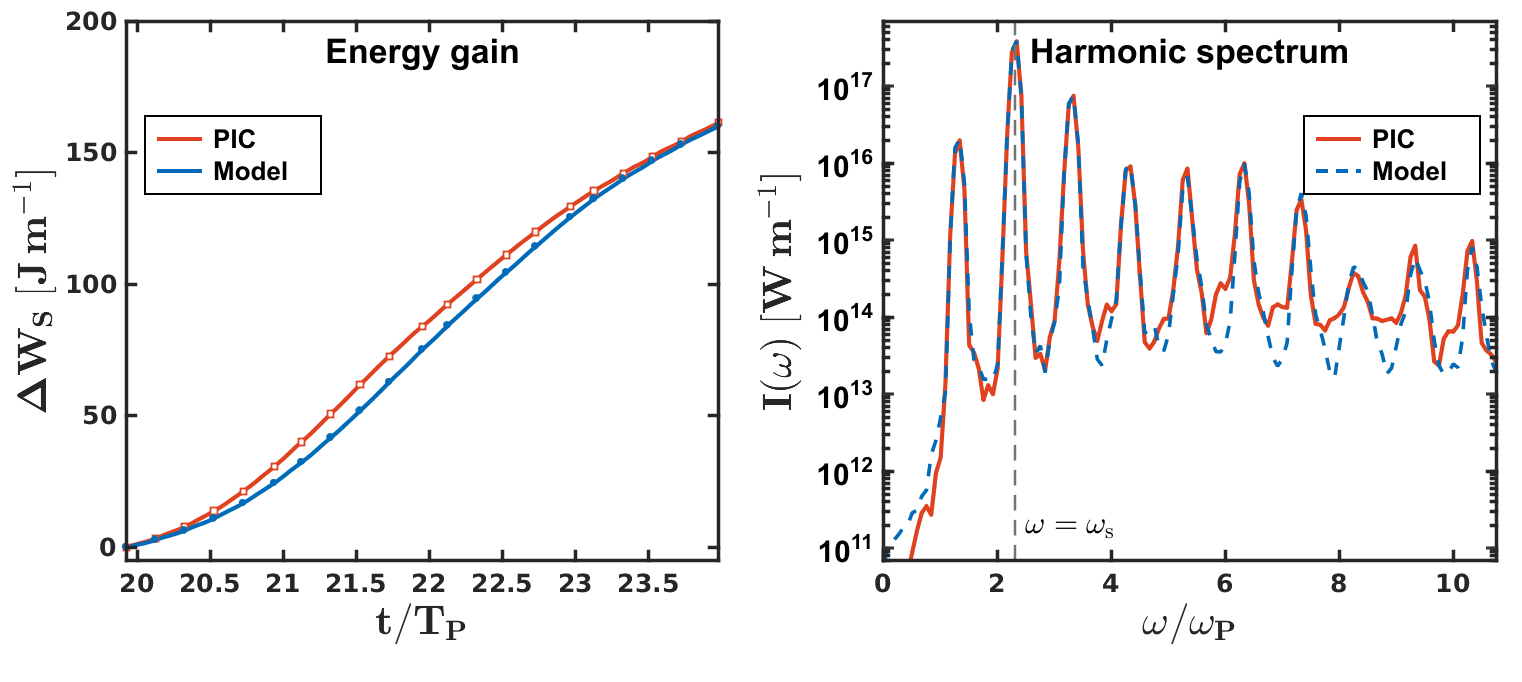}
    \caption{{Comparison between the SPPC prediction and the PIC simulation.} The left panel shows the cumulative energy gain of the s-polarized field during the four dominant modulation periods; the red and blue curves denote the PIC result and the prediction, respectively. The right panel compares the corresponding angle-integrated, time-averaged continuous spectral intensities, with the red solid curve showing the PIC result and the blue dashed curve showing the prediction. The gray vertical dashed line marks the probe frequency $\omega=\omega_S$.}
    \label{fig:energy_gain_spectrum}
\end{figure}

In the PIC simulation, the energy gain is defined as the additional energy carried by the s-polarized field relative to the probe-only simulation under otherwise identical conditions. The left panel of Fig.~\ref{fig:energy_gain_spectrum} compares this PIC result with Eq.~(\ref{eq:cumulative_energy_gain}). The predicted cumulative energy gain closely follows the PIC result throughout the four dominant modulation periods. At the end of the interaction, the total energy gains are
\begin{equation}
    \Delta W_{S,\mathrm{pred}}={160.07~\mathrm{J\,m^{-1}}},
    \quad
    \Delta W_{S,\mathrm{PIC}}={161.41~\mathrm{J\,m^{-1}}},
\end{equation}
corresponding to a relative difference of about ${0.83}\%$. This agreement shows that the local growth associated with the SPPC momentum gap, after accumulation over the interaction region and successive modulation intervals, accounts quantitatively for the net transfer of energy to the s-polarized field.

We next calculate the harmonic spectrum from the reconstructed plasma response. For the $s$-polarized field, the local collisionless scalar approximation gives
\begin{equation}
    \left(\partial_x^2+\partial_y^2\right)E_z
    -\frac{1}{c^2}\partial_t^2E_z
    -\frac{\mu_0 e^2n_e(x,y,t)}{m_e\bar\gamma(x,y,t)}E_z
    =0.
    \label{eq:2d_s_wave}
\end{equation}

{The continuous incident spectrum of the prescribed probe is used as the input, retaining its finite bandwidth. The harmonic-spectrum calculation retains Fourier orders up to $n_{\max}=12$. The frequency-dependent multi-order transfer- and scattering-matrix calculation and its numerical implementation are described in Section~3 of Supplement~1.}

In the right panel of Fig.~\ref{fig:energy_gain_spectrum}, the calculated spectrum reproduces the positions and intensities of the main harmonics at $\omega_S+n\omega_P$. The broader peaks and stronger continuous background in the PIC spectrum can be caused by the finite interaction duration, cycle-to-cycle variation, spatiotemporal inhomogeneity, and off-resonant Floquet coupling.

\begin{figure}[htbp]
    \centering
    \includegraphics[width=0.966\columnwidth,trim={40.00bp 23.00bp 11.00bp 56.00bp},clip]{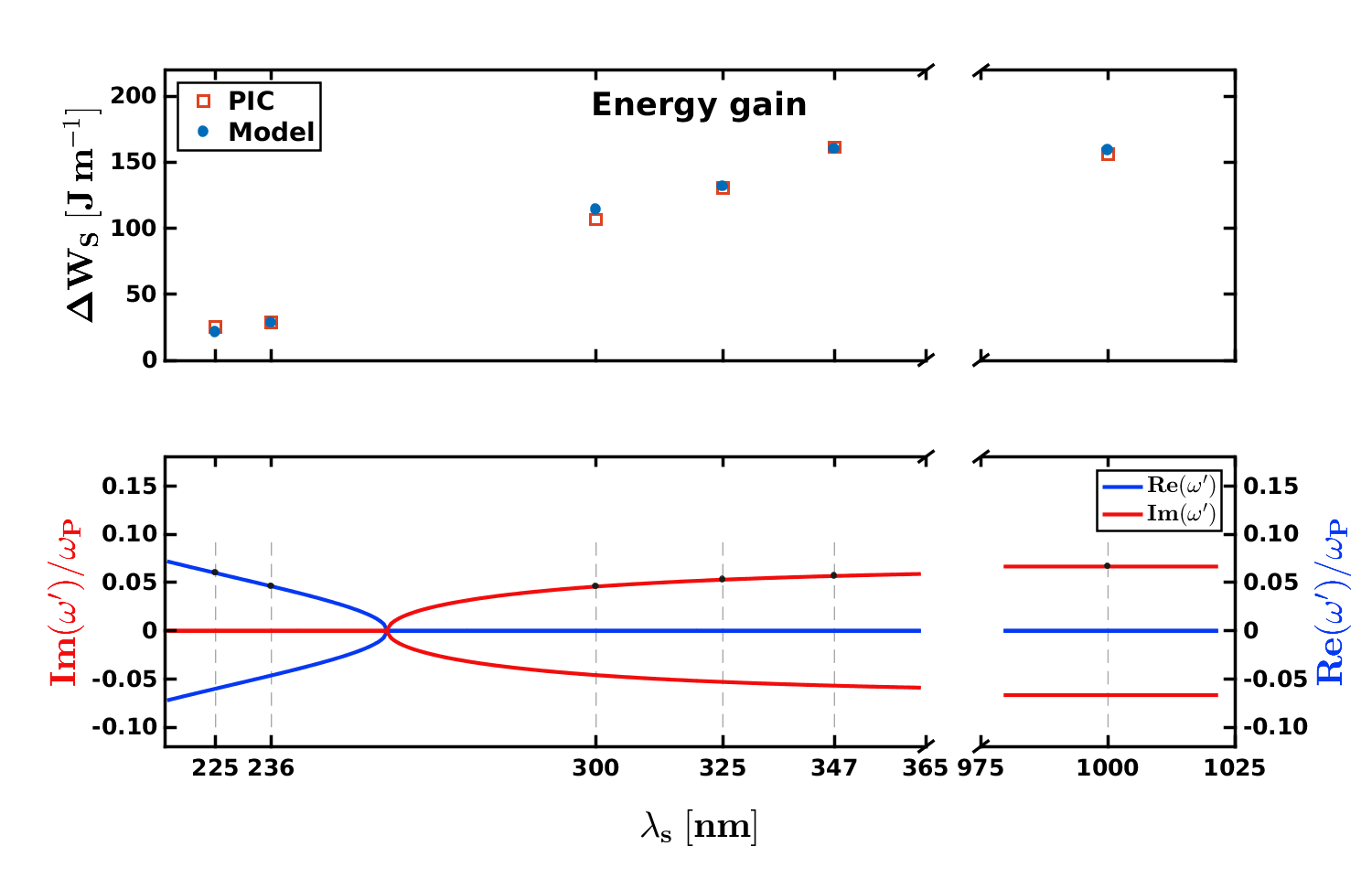}
    \caption{{Energy gain and generalized frequency as functions of the incident probe wavelength. Upper panel: total energy gains of the s-polarized field, with red open squares showing the PIC results and blue filled circles showing the model predictions. Lower panel: real and imaginary parts of the generalized frequency $\omega'$, evaluated for the same representative local modulation as in Fig.~\ref{bandstructure} and shown as functions of the incident probe wavelength $\lambda_S$. The blue curves show $\operatorname{Re}(\omega')/\omega_P$ on the right axis, and the red curves show $\operatorname{Im}(\omega')/\omega_P$ on the left axis. Black dots and gray dashed lines identify the tested probe wavelengths.}}
    \label{fig:probe_wavelength_gain}
\end{figure}

{To examine the dependence of the energy gain on the incident probe frequency, we perform PIC simulations with probe wavelengths of 225, 236, 300, 325, 347, and $1000~\mathrm{nm}$ and compare the gains with the model predictions. The target density plateau is overdense at all six wavelengths. Figure~\ref{fig:probe_wavelength_gain} shows the gain comparison in the upper panel and the real and imaginary parts of $\omega'$ for the same representative local modulation as in Fig.~\ref{bandstructure} in the lower panel.}

{The probes at 225 and $236~\mathrm{nm}$ correspond to an allowed band for the selected local modulation, where $\operatorname{Im}(\omega')=0$, and exhibit relatively small gains. In contrast, the probes at 300, 325, 347, and $1000~\mathrm{nm}$ lie within the momentum bandgap, where the generalized-frequency branches have nonzero imaginary parts, and exhibit substantially larger gains. From 300 to $347~\mathrm{nm}$, the increasing energy gain also follows the increase in the positive imaginary part of the generalized frequency.}

{The generalized frequencies shown here correspond to one representative local region and time interval. The total energy gain is obtained by accumulating the locally evaluated contributions over the entire interaction region and duration, as in Eq.~(\ref{eq:cumulative_energy_gain}), throughout which the local band structures vary. A probe wavelength corresponding to an allowed band for the selected local modulation may therefore fall within a momentum bandgap in other local regions or time intervals and still acquire a finite net gain. For the same reason, the final energy gain is not expected to be directly proportional to the imaginary part of the generalized frequency for this single representative local modulation.}

\section{Extension to the Two-Pump Case}
\label{sec:dual_pump}
We now extend the single-pump configuration in Sec.~\ref{sec:PIC} to two p-polarized pump pulses. The single pump is replaced by two pumps incident symmetrically about the target normal, with signed incidence angles $\theta_{P1}=\theta_P$ and $\theta_{P2}=-\theta_P$, where $\theta_P=\pi/4$. Their wavelengths are $\lambda_{P1}=800~\mathrm{nm}$ and $\lambda_{P2}=1000~\mathrm{nm}$, respectively, and both have the same normalized peak amplitude, $a_{P1}=a_{P2}=2$. The probe and the remaining simulation parameters are unchanged.

{For the dual-pump case, we describe the full plasma response driven by both pumps through Fourier channels labeled by integer pairs $(n_1,n_2)$. A channel supplies frequency and transverse-momentum shifts $n_1\omega_{P1}+n_2\omega_{P2}$ and $n_1k_{P1}\sin\theta_{P1}+n_2k_{P2}\sin\theta_{P2}$ to the probe. These labels also accommodate nonlinear mixed components of the jointly driven response. Combining the allowed shifts with the vacuum dispersion relation gives}
\begin{align}\label{position}
&(k_{(n_1,n_2),x}^L,k_{(n_1,n_2),y}^L)
=\omega_{(n_1,n_2)}(\cos\theta_{(n_1,n_2)},\sin\theta_{(n_1,n_2)}),\notag\\
&\omega_{(n_1,n_2)}
=\omega_S+n_1\omega_{P1}+n_2\omega_{P2},\notag\\
&\sin\theta_{(n_1,n_2)}
=\frac{k_{(n_1,n_2),y}^L}{k_{(n_1,n_2)}^L}
=\frac{n_1\omega_{P1}\sin\theta_{P1}
+n_2\omega_{P2}\sin\theta_{P2}}
{\omega_S+n_1\omega_{P1}+n_2\omega_{P2}}.
\end{align}
\begin{figure}[htbp]
    \centering
    \includegraphics[width=\linewidth]{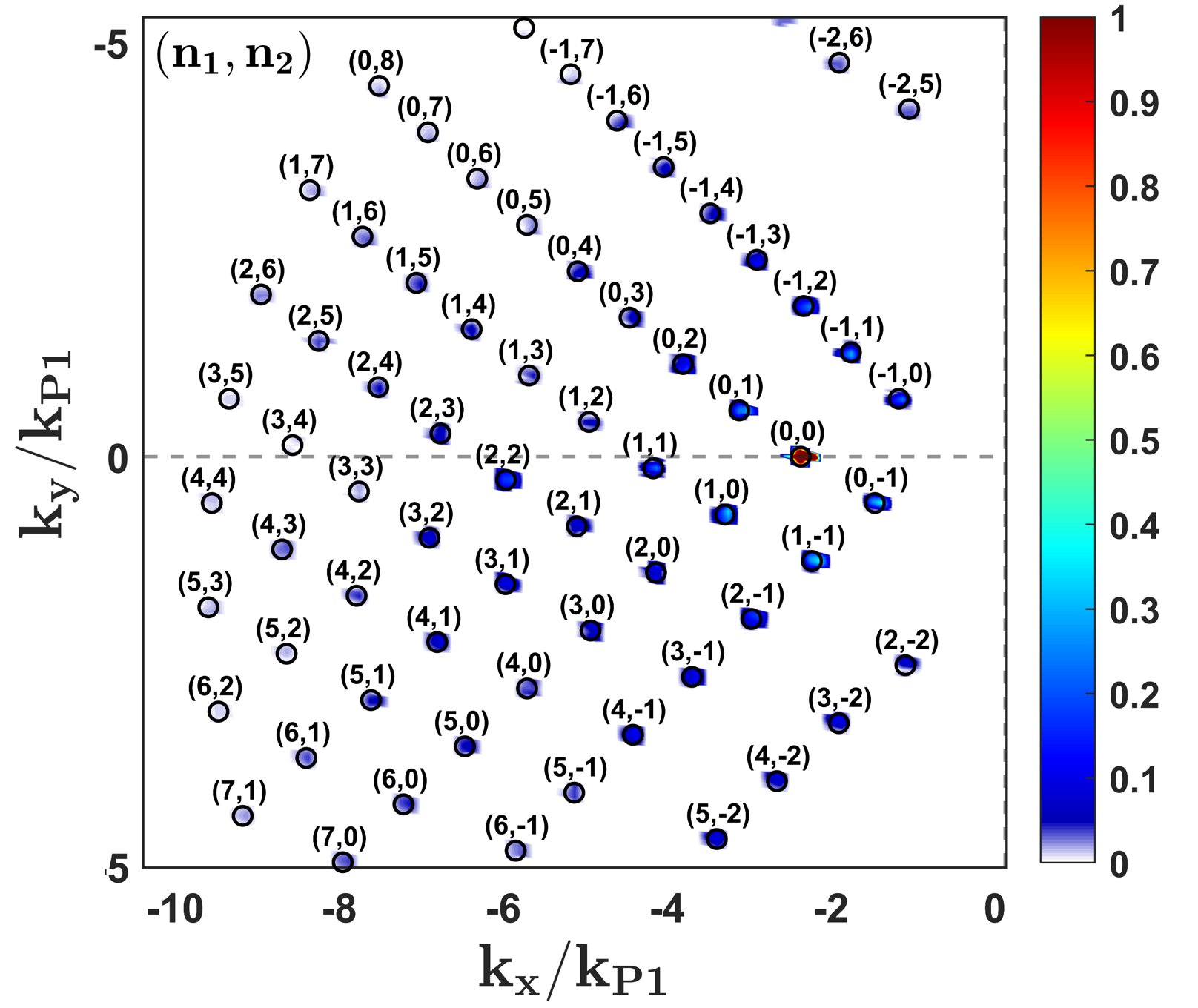}
    \caption{Comparison of the harmonic positions obtained from the dual-pump PIC simulation and the theoretical prediction. The background shows $|E_z(k_x,k_y)|$ in $k$ space, with both axes normalized to $k_{P1}$, and the circles mark the harmonic positions predicted by Eq.~(\ref{position}). {The pairs $(n_1,n_2)$ label the allowed Fourier channels of the jointly driven response.} Since the reflected harmonic components are analyzed, $k_x$ is negative.}
    \label{harmonics_SPC}
\end{figure}

The harmonic positions observed in the dual-pump PIC simulation agree with Eq.~(\ref{position}), including the frequency dependence of the emission angle, as shown in Fig.~\ref{harmonics_SPC}.

\begin{figure}[t]
    \centering
    \includegraphics[width=0.988\columnwidth,trim={0.00bp 33.00bp 16.00bp 7.00bp},clip]{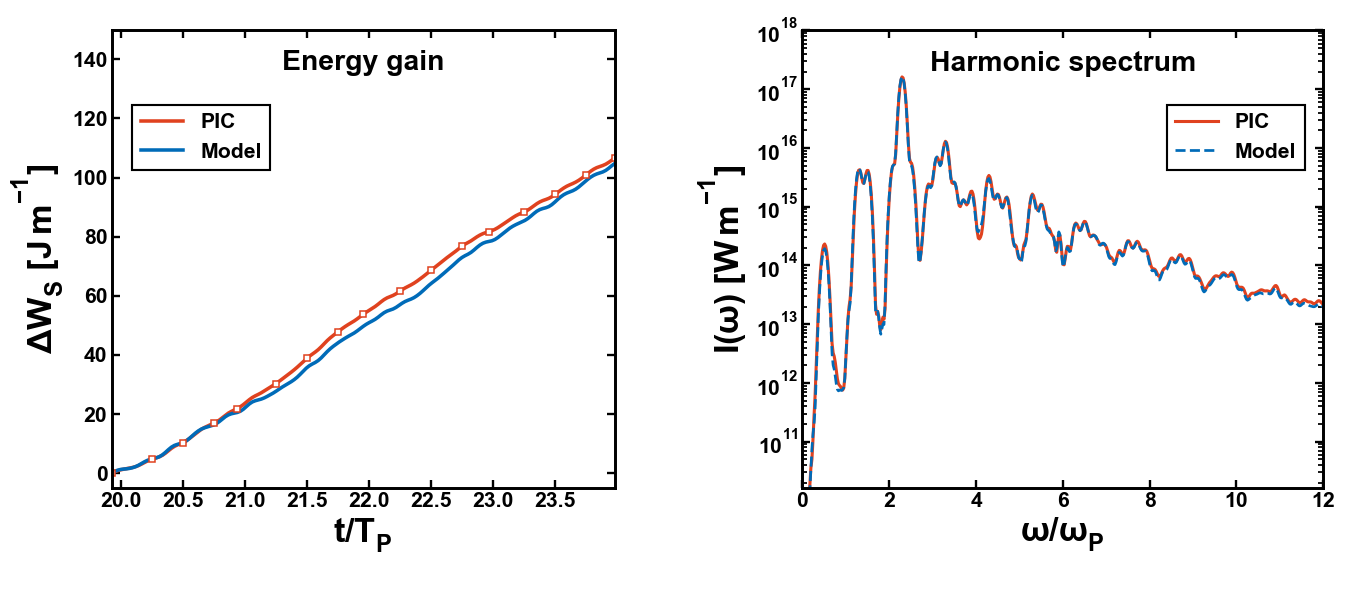}
    \caption{{Comparison of the SPPC prediction with the PIC simulation for the dual-pump configuration. Left panel: cumulative energy gain of the s-polarized field. Right panel: harmonic spectrum of the reflected s-polarized field in absolute intensity. Red curves show the PIC results, and blue curves show the model predictions; the model curve is solid in the left panel and dashed in the right panel. The time and frequency are normalized to the period and frequency of the $800~\mathrm{nm}$ pump, respectively.}}
    \label{fig:dual_pump_gain_spectrum}
\end{figure}

Figure~\ref{fig:dual_pump_gain_spectrum} shows that the model closely follows the PIC gain curve and captures the main spectral features on an absolute intensity scale. The final energy gains are
\begin{equation}
    \Delta W_{S,\mathrm{pred}}=104.73~\mathrm{J\,m^{-1}},
    \quad
    \Delta W_{S,\mathrm{PIC}}=106.51~\mathrm{J\,m^{-1}},
\end{equation}
corresponding to a relative difference of about $1.68\%$.

{Notably, the energy gain in this dual-pump configuration is lower than that in the single-pump case shown in Fig.~\ref{fig:energy_gain_spectrum}, despite the presence of an additional pump. Thus, adding a second pump does not necessarily enhance probe amplification. This result is consistent with treating the plasma modulation as a jointly driven response. Our calculation uses this combined response directly, without assuming that the modulations generated separately by the two pumps can be linearly superposed.}

\begin{figure}[htbp]
    \centering
    \includegraphics[width=0.728\columnwidth,trim={7.80bp 17.64bp 14.76bp 10.44bp},clip]{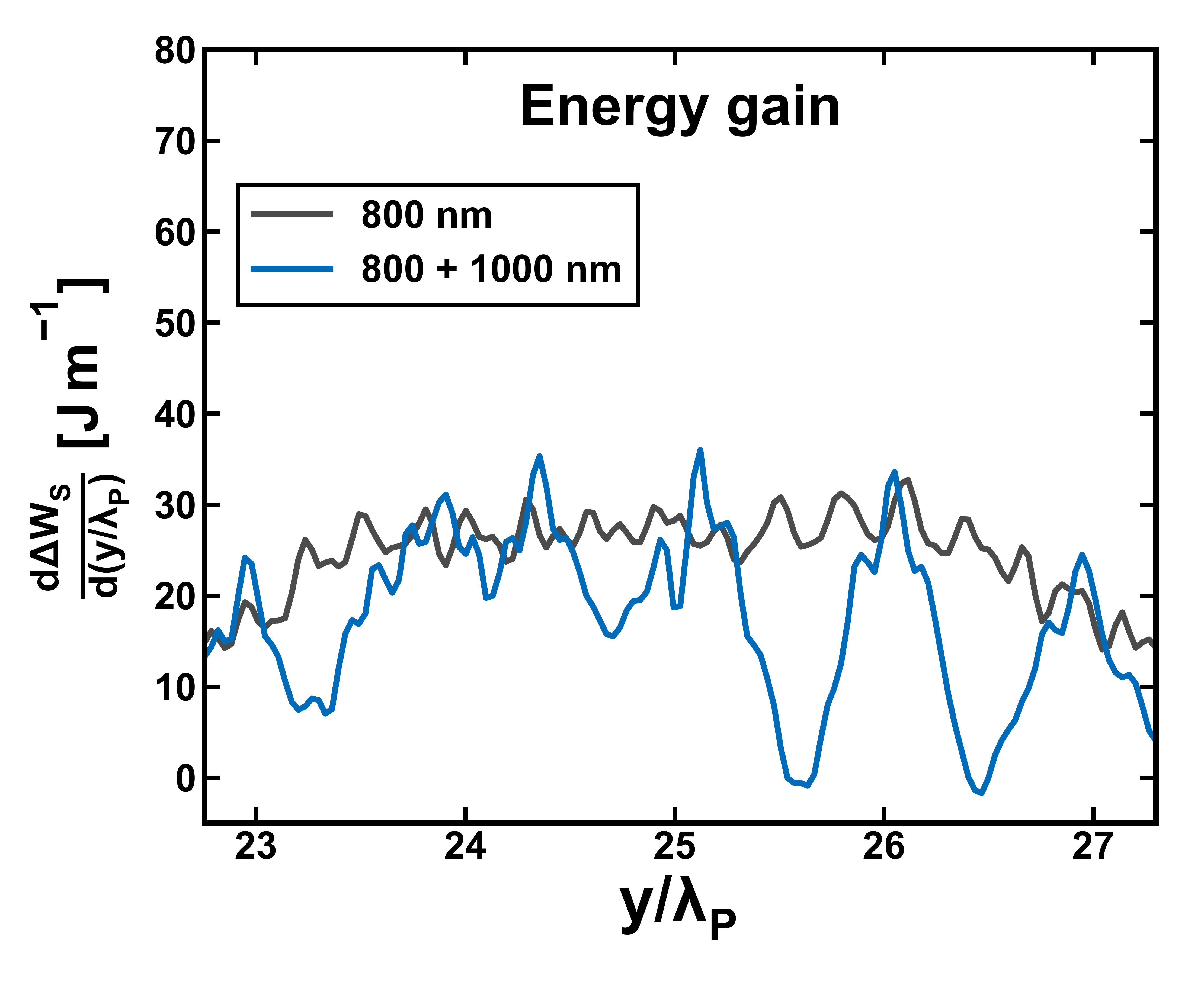}
    \caption{{Transverse profiles of the near-surface probe-energy gain in the single-pump ($800~\mathrm{nm}$, gray) and dual-pump ($800+1000~\mathrm{nm}$, blue) PIC simulations. Here $\lambda_P$ denotes the $800~\mathrm{nm}$ pump wavelength.}}
    \label{fig:single_dual_transverse_gain}
\end{figure}

{To examine the spatial origin of this reduction, Fig.~{\ref{fig:single_dual_transverse_gain}} compares the transverse distributions of the cycle-integrated probe-energy gain in the single- and dual-pump PIC simulations. Both profiles are evaluated over the same near-surface depth and four-cycle time interval. The single-pump profile forms a relatively broad plateau, whereas the dual-pump profile exhibits alternating enhanced-gain regions and pronounced low-gain troughs. In several regions, the dual-pump gain exceeds the single-pump value, but these local enhancements do not compensate for the losses in the suppressed regions.}

{The two symmetrically incident pumps produce a periodic interference pattern in the surface-normal driving field $E_x$. Where their $E_x$ components interfere constructively, the enhanced normal drive can strengthen the periodic electron-layer motion and the associated effective-permittivity modulation, thereby increasing local amplification. This is consistent with the portions of the dual-pump curve that exceed the single-pump result in Fig.~{\ref{fig:single_dual_transverse_gain}}. Conversely, destructive interference weakens the normal drive and the corresponding effective modulation, reducing the local amplification. The resulting contrast provides a physical interpretation of the lower spatially integrated gain despite stronger amplification in some regions.}

{Because the two pumps have different frequencies, this interference pattern drifts along the surface at $v_b=\Delta\omega_P/\Delta k_{P,y}$, where $\Delta\omega_P=\omega_{P1}-\omega_{P2}$ and $\Delta k_{P,y}=k_{P1}\sin\theta_{P1}-k_{P2}\sin\theta_{P2}$. This drift is much slower than the phase propagation of the pump-driven surface modulation in the single-pump case. Consequently, averaging over the few cycles considered here does not fully wash out the transverse contrast between enhanced- and suppressed-driving regions; the contrast remains visible in the cycle-integrated gain profile in Fig.~{\ref{fig:single_dual_transverse_gain}}.}

\section{Conclusions}
In conclusion, we have shown that a relativistically driven plasma surface acts as a spatiotemporal plasma photonic crystal (SPPC) for a weak probe. The spatial and temporal periodicity of the SPPC determines the allowed harmonic frequencies and emission angles. For two pumps, the same framework identifies the mixed frequency--momentum channels of the jointly driven plasma response. The predicted harmonic positions agree with the PIC results in both configurations.

The SPPC picture explains how the probe draws energy from the pump-driven modulation. We use the complex Floquet bands to quantify the probe energy gain and the same plasma response to predict the harmonic spectrum. The final energy gains differ from the PIC results by 0.83\% and 1.68\% in the single- and dual-pump cases, respectively. The predicted harmonic spectra also agree well with the PIC spectra in both cases.

Multiple pumps offer a route to reconfigurable SPPCs. Varying their frequencies, incidence angles, relative phases, and polarizations could tailor the plasma modulation and thereby control harmonic frequencies, emission directions, and conversion efficiency. Structured drivers could extend this control to the harmonic wavefront. For example, Laguerre--Gaussian beams could imprint azimuthal structure on the plasma modulation, offering a route to attosecond vortex radiation. More broadly, treating periodically driven plasmas as spatiotemporal photonic crystals provides a useful framework for understanding and designing new regimes of laser-plasma interaction.

\section*{Acknowledgments}
This work was supported by the National Natural Science Foundation of China
(Grant Nos. 12225505, 12475245, and 12135009), the Fundamental and Interdisciplinary Disciplines
Breakthrough Plan of the Ministry of Education of China
(JYB2025XDXM204), the Shanghai Jiao Tong University 2030 Initiative,
and, in part, by the State Key Laboratory of Dark Matter Physics.
The computations in this paper were performed on the $\pi$ 2.0 cluster
supported by the Center for High Performance Computing at Shanghai Jiao Tong University.

\begin{backmatter}
\bmsection{Supplemental document}
See Supplement~1 for supporting content.
\end{backmatter}

\bibliography{references}

\begin{thebibliography}{10}
\newcommand{\enquote}[1]{``#1''}

\bibitem{Sharabi:22}
Y.~Sharabi, A.~Dikopoltsev, E.~Lustig, \emph{et~al.}, \enquote{Spatiotemporal
  photonic crystals,} {\protect\JournalTitle{Optica}} \textbf{9}, 585--592
  (2022).

\bibitem{Ye2025ActiveMetasurface}
X.~Ye, Y.~G. Wang, J.~F. Yao, \emph{et~al.}, \enquote{Realization of
  spatiotemporal photonic crystals based on active metasurface,}
  {\protect\JournalTitle{Laser \& Photonics Reviews}} \textbf{19}, 2401345
  (2025).

\bibitem{PhysRevLett.58.2059}
E.~Yablonovitch, \enquote{Inhibited spontaneous emission in solid-state physics
  and electronics,} {\protect\JournalTitle{Physical Review Letters}}
  \textbf{58}, 2059--2062 (1987).

\bibitem{PhysRevLett.58.2486}
S.~John, \enquote{Strong localization of photons in certain disordered
  dielectric superlattices,} {\protect\JournalTitle{Physical Review Letters}}
  \textbf{58}, 2486--2489 (1987).

\bibitem{Lustig:18}
E.~Lustig, Y.~Sharabi, and M.~Segev, \enquote{Topological aspects of photonic
  time crystals,} {\protect\JournalTitle{Optica}} \textbf{5}, 1390--1395
  (2018).

\bibitem{PhysRevE.75.046607}
F.~Biancalana, A.~Amann, A.~V. Uskov, and E.~P. O'Reilly, \enquote{Dynamics of
  light propagation in spatiotemporal dielectric structures,}
  {\protect\JournalTitle{Phys. Rev. E}} \textbf{75}, 046607 (2007).

\bibitem{doi:10.1126/science.abo3324}
M.~Lyubarov, Y.~Lumer, A.~Dikopoltsev, \emph{et~al.}, \enquote{Amplified
  emission and lasing in photonic time crystals,}
  {\protect\JournalTitle{Science}} \textbf{377}, 425--428 (2022).

\bibitem{Wang2023MetasurfacePTC}
X.~Wang, M.~S. Mirmoosa, V.~S. Asadchy, \emph{et~al.},
  \enquote{Metasurface-based realization of photonic time crystals,}
  {\protect\JournalTitle{Science Advances}} \textbf{9}, eadg7541 (2023).

\bibitem{PhysRevLett.76.50}
P.~Gibbon, \enquote{Harmonic generation by femtosecond laser-solid interaction:
  A coherent ``water-window'' light source?} {\protect\JournalTitle{Phys. Rev.
  Lett.}} \textbf{76}, 50--53 (1996).

\bibitem{Timmis2026EfficiencyOptimized}
R.~J.~L. Timmis, C.~R.~J. Fitzpatrick, J.~P. Kennedy, \emph{et~al.},
  \enquote{Efficiency-optimized relativistic plasma harmonics for extreme
  fields,} {\protect\JournalTitle{Nature}} \textbf{652}, 1153--1158 (2026).

\bibitem{10.1063/1.870766}
S.~V. Bulanov, N.~M. Naumova, and F.~Pegoraro, \enquote{{Interaction of an
  ultrashort, relativistically strong laser pulse with an overdense plasma},}
  {\protect\JournalTitle{Physics of Plasmas}} \textbf{1}, 745--757 (1994).

\bibitem{Lichters_1996}
R.~Lichters, J.~Meyer-ter Vehn, and A.~Pukhov, \enquote{Short-pulse laser
  harmonics from oscillating plasma surfaces driven at relativistic intensity,}
  {\protect\JournalTitle{Physics of Plasmas}} \textbf{3}, 3425--3437 (1996).

\bibitem{Dromey2006RelativisticLimit}
B.~Dromey, M.~Zepf, A.~Gopal, \emph{et~al.}, \enquote{High harmonic generation
  in the relativistic limit,} {\protect\JournalTitle{Nature Physics}}
  \textbf{2}, 456--459 (2006).

\bibitem{PhysRevE.74.046404}
T.~Baeva, S.~Gordienko, and A.~Pukhov, \enquote{Theory of high-order harmonic
  generation in relativistic laser interaction with overdense plasma,}
  {\protect\JournalTitle{Phys. Rev. E}} \textbf{74}, 046404 (2006).

\bibitem{Xia2023AngularDispersed}
Y.~Xia, B.~Li, M.~Chen, \emph{et~al.}, \enquote{Efficient angular dispersed
  high-order harmonic generation by dual laser pulses interacting with a solid
  target,} {\protect\JournalTitle{Physics of Plasmas}} \textbf{30}, 123108
  (2023).

\bibitem{Wang2025IsolatedAttosecond}
X.~Wang, J.~Gao, F.~Liu, \emph{et~al.}, \enquote{Isolated attosecond pulse from
  a solid target driven by two laser pulses,} {\protect\JournalTitle{Physics of
  Plasmas}} \textbf{32}, 083303 (2025).

\bibitem{Zhang2026OPGPRR}
H.-J. Zhang, B.-Y. Li, X.-Y. An, \emph{et~al.}, \enquote{Generation of
  attosecond pulses from relativistic oscillating plasma gratings,}
  {\protect\JournalTitle{Physical Review Research}} \textbf{8}, 023261 (2026).

\bibitem{arber2015contemporary}
T.~D. Arber, K.~Bennett, C.~S. Brady, \emph{et~al.}, \enquote{Contemporary
  particle-in-cell approach to laser-plasma modelling,}
  {\protect\JournalTitle{Plasma Physics and Controlled Fusion}} \textbf{57},
  113001 (2015).

\bibitem{PhysRevE.100.053207}
B.~Y. Li, F.~Liu, M.~Chen, \emph{et~al.}, \enquote{High-quality high-order
  harmonic generation through preplasma truncation,}
  {\protect\JournalTitle{Phys. Rev. E}} \textbf{100}, 053207 (2019).

\bibitem{li2022experimental}
B.~Y. Li, F.~Liu, M.~Chen, \emph{et~al.}, \enquote{Experimental demonstration
  of efficient harmonic generation via surface plasma compression with lasers,}
  {\protect\JournalTitle{Physical Review Letters}} \textbf{128}, 244801 (2022).

\bibitem{PhysRevB.68.045121}
S.~Mishra and S.~Satpathy, \enquote{One-dimensional photonic crystal: The
  kronig-penney model,} {\protect\JournalTitle{Phys. Rev. B}} \textbf{68},
  045121 (2003).

\end{thebibliography}


\begin{thebibliography}{1}
\newcommand{\enquote}[1]{``#1''}

\bibitem{PhysRevE.100.053207}
B.~Y. Li, F.~Liu, M.~Chen, \emph{et~al.}, \enquote{High-quality high-order
  harmonic generation through preplasma truncation,}
  {\protect\JournalTitle{Phys. Rev. E}} \textbf{100}, 053207 (2019).

\bibitem{li2022experimental}
B.~Y. Li, F.~Liu, M.~Chen, \emph{et~al.}, \enquote{Experimental demonstration
  of efficient harmonic generation via surface plasma compression with lasers,}
  {\protect\JournalTitle{Physical Review Letters}} \textbf{128}, 244801 (2022).

\end{thebibliography}

\end{document}


\maketitle

\section{Particle-in-Cell Simulation Parameters}
\label{sec:supp-pic-parameters}

\paragraph{Laser pulses.}
The $p$-polarized pump ($\lambda_P=800~\mathrm{nm}$, $a_P=2$) is incident at $\theta_P=\pi/4$. The $s$-polarized probe ($\lambda_S=347~\mathrm{nm}$, $a_S=0.5$) is normally incident. Their focal spot sizes are $3\lambda_P$ and $8\lambda_P$, respectively, and $a_j=eE_j/(m_ec\omega_j)$ for $j=P,S$. Each pulse has a $\sin^2(\pi t/2\tau_j)$ envelope, with $\tau_P=10T_P$, $\tau_S=5T_P$, and $T_P=2\pi/\omega_P$. The delay is set so that the intensity maxima reach the target simultaneously.

\paragraph{Target.}
The solid target is $2\lambda_P$ thick with a plateau electron density of $20n_c$, where $n_c=\varepsilon_0m_e\omega_P^2/e^2=1.7\times10^{21}~\mathrm{cm}^{-3}$. The plateau begins at $x_0=8\lambda_P$. The preplasma follows $n_e(x)=20n_c\exp[(x-x_0)/d]$ for $x<x_0$, with $d=0.1\lambda_P$, and is truncated at $7n_c\cos\theta_P$. This choice is motivated by studies of preplasma truncation \cite{PhysRevE.100.053207} and laser-driven surface plasma compression \cite{li2022experimental} for efficient harmonic generation.

\paragraph{Numerical setup.}
The two-dimensional EPOCH simulations use a $12\lambda_P\times50\lambda_P$ box (along $x$ and $y$), with $\Delta x=\Delta y=\lambda_P/250$ and 64 electron and 16 ion macroparticles per cell. The pulses enter through the left boundary at their prescribed angles. The other boundaries are open. For the two-pump configuration, only the pump settings specified in Sec.~4 of the main text change.

\section{Wave Equation in the Rotated Coordinates}
\label{app:rotated-wave-equation}

Within the effective-medium description, Maxwell's equations for the probe field are
\begin{align}\label{maxwell}
\nabla\times E+\frac{\partial B}{\partial t}&=0,\\
\nabla\times H-\frac{\partial D}{\partial t}&=0,
\end{align}
{For the frequency-selection derivation, we neglect the weak probe feedback on the pump-driven plasma and evaluate the absolute permittivity $\varepsilon(t,x)=\varepsilon_0\varepsilon_{\mathrm{eff}}(x,y_0,t;\omega_S)$ at a fixed transverse position $y_0$. The instantaneous constitutive relations are}
\begin{align}
   D&=\varepsilon(t,x)E,\label{eq:instantaneous_constitutive}\\
   H&=B/\mu_0.
\end{align}
In the geometrically rotated coordinate system, the permittivity depends only on $x'$. Because the pump-driven plasma response approximately recurs from one optical cycle to the next, $\varepsilon'(x')$ can be treated as periodic over the relevant interaction interval. We denote the period in the rotated coordinate by $l'$,
\begin{equation}
    \varepsilon'(x')=\varepsilon'(x'+l').
\end{equation}
For a modulation frequency \(\omega_P\), the temporal period in the original coordinates is \(T_P=2\pi/\omega_P\). Under the rotation \(x'=-t\sin\theta+x\cos\theta\), a temporal translation \(\Delta t=T_P\) at fixed \(x\) corresponds to \(\Delta x'=-T_P\sin\theta\). Since the permittivity is independent of \(t'\), its period along \(x'\) is therefore \(l'=|\Delta x'|=2\pi|\sin\theta|/\omega_P\).

{Expressing the permittivity in the rotated coordinates, we write the constitutive relation as}
\begin{equation}\label{eq:rotated_constitutive}
    {D=\varepsilon'(x')E.}
\end{equation}

Using Eqs.~(14) and (17) of the main text, the coordinate derivatives and fields in the original coordinates are given by
\begin{align}
    \frac{\partial}{\partial t}&=\cos\theta \frac{\partial}{\partial t'}-\sin\theta \frac{\partial}{\partial x'},\notag\\
    \frac{\partial}{\partial x}&=\cos\theta \frac{\partial}{\partial x'}+\sin\theta \frac{\partial}{\partial t'}.
\end{align}
The field components transform as
\begin{align}
    E_z&=\cos\theta E_z'-\sin\theta B_y',\notag\\
    B_y&=\cos\theta B_y'+\sin\theta E_z'.
\end{align}
For the one-dimensional geometry considered here, the other field components and derivatives remain unchanged.

Substituting these relations into Faraday's law gives
\begin{align}
    &\left(\nabla\times E+\frac{\partial B}{\partial t}\right)_y
    =-\frac{\partial E_z}{\partial x}
    +\frac{\partial B_y}{\partial t}\notag\\
    &=-
    \left(
    \cos\theta\frac{\partial}{\partial x'}
    +\sin\theta\frac{\partial}{\partial t'}
    \right)
    \left(
    \cos\theta E_z'-\sin\theta B_y'
    \right)\notag\\
    &\quad+
    \left(
    \cos\theta\frac{\partial}{\partial t'}
    -\sin\theta\frac{\partial}{\partial x'}
    \right)
    \left(
    \cos\theta B_y'+\sin\theta E_z'
    \right)\notag\\
    &=-\frac{\partial E_z'}{\partial x'}
    +\frac{\partial B_y'}{\partial t'}=0.
    \label{maxwell1}
\end{align}
Thus, this part of Maxwell's equations retains its form under the geometric rotation. The remaining derivatives are
\begin{align}
    &\frac{\partial B_y}{\partial x}
    =
    \left(
    \cos\theta\frac{\partial}{\partial x'}
    +\sin\theta\frac{\partial}{\partial t'}
    \right)
    \left(
    \cos\theta B_y'+\sin\theta E_z'
    \right)\notag\\
    &=\cos^2\theta\frac{\partial B_y'}{\partial x'}
    +\sin\theta\cos\theta
    \left(
    \frac{\partial E_z'}{\partial x'}
    +\frac{\partial B_y'}{\partial t'}
    \right)
    +\sin^2\theta\frac{\partial E_z'}{\partial t'},
\end{align}
and
  \begin{align}
    &  \frac{\partial D_z}{\partial t}
  =
    \left(
    \cos\theta\frac{\partial}{\partial t'}
    -\sin\theta\frac{\partial}{\partial x'}
    \right)
    \left[
    \varepsilon'
    \left(
    \cos\theta E_z'-\sin\theta B_y'
    \right)
    \right]\notag\\
    &=\cos^2\theta\varepsilon'
    \frac{\partial E_z'}{\partial t'}
    -\sin\theta\cos\theta\varepsilon'
    \frac{\partial B_y'}{\partial t'}
    -\sin\theta\cos\theta\varepsilon'
    \frac{\partial E_z'}{\partial x'}\notag\\
    &\quad+\sin^2\theta\varepsilon'
    \frac{\partial B_y'}{\partial x'}
    -\sin\theta\cos\theta E_z'
    \frac{\partial\varepsilon'}{\partial x'}
    +\sin^2\theta B_y'
    \frac{\partial\varepsilon'}{\partial x'}.
\end{align}
Taking the derivative with respect to \(t'\) and using Eq.~(\ref{maxwell1}), we obtain
\begin{align}
    &\frac{\partial}{\partial t'}
    \left(
    \frac{\partial B_y}{\partial x}
    -\mu_0\frac{\partial D_z}{\partial t}
    \right)\notag\\
    &=
    \left(
    \cos^2\theta-\mu_0\varepsilon'\sin^2\theta
    \right)
    \frac{\partial^2E_z'}{\partial x'^2}
    +
    \left(
    \sin^2\theta-\mu_0\varepsilon'\cos^2\theta
    \right)
    \frac{\partial^2E_z'}{\partial t'^2}\notag\\
    &\quad+
    2\sin\theta\cos\theta
    \left(
    1+\mu_0\varepsilon'
    \right)
    \frac{\partial^2E_z'}{\partial t'\partial x'}\notag\\
    &\quad+
    \mu_0
    \left(
    \sin\theta\cos\theta\frac{\partial E_z'}{\partial t'}
    -\sin^2\theta\frac{\partial E_z'}{\partial x'}
    \right)
    \frac{\partial\varepsilon'}{\partial x'}=0.
\end{align}

Taking
\[
E_z'(x',t')=E_{z0}'(x')e^{-i\omega't'},
\quad
B_y'(x',t')=B_{y0}'(x')e^{-i\omega't'},
\]
the electric-field amplitude satisfies
\begin{align}
    &\frac{\mathrm{d}^2E_{z0}'}{\mathrm{d}x'^2}
    -
    \frac{
    \mu_0\sin^2\theta\,\partial_{x'}\varepsilon'
    +2i\omega'\sin\theta\cos\theta
    \left(1+\mu_0\varepsilon'\right)
    }{
    \cos^2\theta-\mu_0\varepsilon'\sin^2\theta
    }
    \frac{\mathrm{d}E_{z0}'}{\mathrm{d}x'}\notag\\
    &+
    \frac{
    \left(
    \mu_0\varepsilon'\cos^2\theta-\sin^2\theta
    \right)\omega'^2
    -i\omega'\mu_0\sin\theta\cos\theta\,
    \partial_{x'}\varepsilon'
    }{
    \cos^2\theta-\mu_0\varepsilon'\sin^2\theta
    }
    E_{z0}'=0.\label{helmhotzeq}
\end{align}

\section{Multi-Order Floquet Transfer-Matrix Method}
\label{app:floquet-transfer-matrix}

We write the two-dimensional wave equation for the probe field [Eq.~(41) of the main text] in terms of the effective plasma frequency,
\begin{equation}
    {\left(\partial_x^2+\partial_y^2\right)E_z
    -\frac{1}{c^2}\partial_t^2E_z
    -\frac{\omega_{pe,\mathrm{eff}}^2(x,y,t)}{c^2}E_z
    =0.}
    \label{eq:local_2d_s_wave}
\end{equation}
The electron density $n_e$ and mean Lorentz factor $\bar\gamma$ from the pump-only PIC simulation determine $\omega_{pe,\mathrm{eff}}^2(x,y,t)=e^2n_e(x,y,t)/[\varepsilon_0m_e\bar\gamma(x,y,t)]$. The prescribed incident probe field sets the boundary condition.

{For a time-independent plasma response, this equation recovers $\varepsilon_{\mathrm{eff}}(\omega)=1-\omega_{pe,\mathrm{eff}}^2/\omega^2$. The continuous incident spectrum is evaluated frequency by frequency. For each incident frequency $\omega$, we fix $y=y_j$, define $E_j(x,t;\omega)\equiv E_z(x,y_j,t;\omega)$ and $\omega_{pe,\mathrm{eff},j}^2(x,t)\equiv\omega_{pe,\mathrm{eff}}^2(x,y_j,t)$, and expand the plasma response and field in temporal Fourier components:}
\begin{equation}
    {\begin{aligned}
    \omega_{pe,\mathrm{eff},j}^2(x,t)
    &=\sum_l\left(\omega_{pe,\mathrm{eff}}^2\right)_{l,j}(x)e^{-il\omega_Pt},\\
    E_j(x,t;\omega)
    &=\sum_m E_{m,j}(x;\omega)e^{-i\omega_mt},
    \qquad \omega_m=\omega+m\omega_P.
    \end{aligned}}
\end{equation}
{The transverse momentum selection rule gives $k_{m,y}=m k_P\sin\theta_P$. Retaining the transverse derivative through $\partial_y^2E_m=-k_{m,y}^2E_m$ in Eq.~(\ref{eq:local_2d_s_wave}) then gives the coupled Floquet equations}
\begin{equation}
    \begin{aligned}
    &\frac{\mathrm{d}^2E_{m,j}}{\mathrm{d}x^2}\\
    &\quad+\sum_n\left[
    \left(\frac{\omega_m^2}{c^2}-k_{m,y}^2\right)\delta_{mn}
    -\frac{\left(\omega_{pe,\mathrm{eff}}^2\right)_{m-n,j}(x)}{c^2}
    \right]E_{n,j}=0.
    \end{aligned}
\end{equation}

Within each spatial layer, the Fourier coefficients are treated as locally constant. After truncating the expansion to $m=-M,\ldots,M$, we introduce the Floquet-amplitude vector $\boldsymbol{E}_j=(E_{-M,j},\ldots,E_{M,j})^{\mathrm T}$. {For each incident frequency $\omega$ and local element $(x_i,y_j,t_p)$,} the coupled equations become
\begin{equation}
    \begin{aligned}
    \frac{\mathrm{d}^2\boldsymbol{E}_j}{\mathrm{d}x^2}
    +\boldsymbol{C}_{ijp}{(\omega)}\boldsymbol{E}_j&=0,\\
    \left(\boldsymbol{C}_{ijp}{(\omega)}\right)_{mn}
    &{=\left(\frac{\omega_m^2}{c^2}-k_{m,y}^2\right)\delta_{mn}
    -\frac{\left(\omega_{pe,\mathrm{eff}}^2\right)_{m-n}^{(ijp)}}{c^2}.}
    \end{aligned}
\end{equation}
{The dependence of the subsequent transfer and scattering matrices on the incident frequency $\omega$ is suppressed for brevity.}
Defining $\boldsymbol{\Psi}_j=(\boldsymbol{E}_j,\partial_x\boldsymbol{E}_j)^{\mathrm T}$ converts this equation into first-order form,
\begin{equation}
    \frac{\mathrm{d}\boldsymbol{\Psi}_j}{\mathrm{d}x}
    =\boldsymbol{A}_{ijp}\boldsymbol{\Psi}_j,
    \quad
    \boldsymbol{A}_{ijp}
    =
    \begin{pmatrix}
        \boldsymbol{0} & \boldsymbol{I}\\
        -\boldsymbol{C}_{ijp} & \boldsymbol{0}
    \end{pmatrix}.
\end{equation}
The state vector across a layer of thickness $\Delta x_i$ is therefore propagated by
\begin{equation}
    \boldsymbol{\Psi}_j(x+\Delta x_i)
    =\boldsymbol{P}_{ijp}\boldsymbol{\Psi}_j(x),
    \quad
    \boldsymbol{P}_{ijp}
    =\exp\!\left(\boldsymbol{A}_{ijp}\Delta x_i\right).
\end{equation}

For fixed $(j,p)$, the field traverses the $N_x$ layers successively along the surface-normal direction. The propagator across the complete skin-depth region is therefore the ordered product
\begin{equation}
    \boldsymbol{P}_{jp}
    =\overleftarrow{\prod_{i=1}^{N_x}}
    \boldsymbol{P}_{ijp},
\end{equation}
where the arrow denotes surface-normal ordering, with matrices of larger $i$ acting from the left.

To separate waves propagating in the $+x$ and $-x$ directions, we write
\begin{equation}
    \boldsymbol{\Psi}_j
    =\boldsymbol{Q}
    \begin{pmatrix}
        \boldsymbol{E}^{+}\\
        \boldsymbol{E}^{-}
    \end{pmatrix},
\end{equation}
where $\boldsymbol{Q}$ contains the corresponding Floquet eigenmodes. The transfer matrix in this modal basis is
\begin{equation}
    \boldsymbol{T}_{jp}
    =\boldsymbol{Q}_{R}^{-1}
    \boldsymbol{P}_{jp}
    \boldsymbol{Q}_{L}
    =
    \begin{pmatrix}
        \boldsymbol{T}_{11} & \boldsymbol{T}_{12}\\
        \boldsymbol{T}_{21} & \boldsymbol{T}_{22}
    \end{pmatrix},
\end{equation}
and thus
\begin{equation}
    \begin{pmatrix}
        \boldsymbol{E}_{R}^{+}\\
        \boldsymbol{E}_{R}^{-}
    \end{pmatrix}
    =\boldsymbol{T}_{jp}
    \begin{pmatrix}
        \boldsymbol{E}_{L}^{+}\\
        \boldsymbol{E}_{L}^{-}
    \end{pmatrix}.
\end{equation}
Here $(\boldsymbol{E}_{L}^{+},\boldsymbol{E}_{R}^{-})$ are the incident amplitudes, whereas $(\boldsymbol{E}_{L}^{-},\boldsymbol{E}_{R}^{+})$ are the outgoing amplitudes. Assuming that $\boldsymbol{T}_{22}$ is nonsingular, solving the second block row for $\boldsymbol{E}_{L}^{-}$ and substituting it into the first gives
\begin{equation}
    \begin{pmatrix}
        \boldsymbol{E}_{L}^{-}\\
        \boldsymbol{E}_{R}^{+}
    \end{pmatrix}
    =\boldsymbol{S}_{jp}
    \begin{pmatrix}
        \boldsymbol{E}_{L}^{+}\\
        \boldsymbol{E}_{R}^{-}
    \end{pmatrix},
\end{equation}
with
\begin{equation}
\begin{aligned}
    (\boldsymbol{S}_{jp})_{11}
    &=-\boldsymbol{T}_{22}^{-1}\boldsymbol{T}_{21},
    &
    (\boldsymbol{S}_{jp})_{12}
    &=\boldsymbol{T}_{22}^{-1},\\
    (\boldsymbol{S}_{jp})_{21}
    &=\boldsymbol{T}_{11}
      -\boldsymbol{T}_{12}\boldsymbol{T}_{22}^{-1}\boldsymbol{T}_{21},
    &
    (\boldsymbol{S}_{jp})_{22}
    &=\boldsymbol{T}_{12}\boldsymbol{T}_{22}^{-1}.
\end{aligned}
\end{equation}
{For each $(j,p)$ and incident frequency $\omega$, this scattering matrix maps the probe field onto reflected and transmitted Floquet sidebands with transverse wave vectors $k_{m,y}=m k_{P,y}$. The continuous input frequencies account for the finite probe bandwidth, while the integer channel order determines the transverse momentum shift. The outgoing amplitudes are assembled at their corresponding output frequencies, and the angle-integrated spectrum is obtained by summing the channel intensities over the transverse subregions $j$ and selected modulation intervals $p$. The calculation thus retains the order-dependent transverse wave vector, although the angle-integrated plot does not separately display the individual transverse-momentum channels.} The simultaneous agreement of the predicted cumulative energy gain and angle-integrated spectrum with the PIC results shows that the reduced model captures the dominant conversion process and provides quantitative evidence for the SPPC amplification mechanism.

\bibliography{references}